\documentclass[longauth]{aa}
\usepackage[utf8]{inputenc}
\usepackage[varg]{txfonts}
\usepackage[usenames, dvipsnames]{color}
\usepackage{amsmath}
\usepackage{graphicx}
\usepackage[english]{babel}
\usepackage{siunitx}
\usepackage{float}
\usepackage{url}
\usepackage{longtable}
\usepackage{fancyhdr}
\usepackage{natbib}
\usepackage[normalem]{ulem}
\usepackage[version=4]{mhchem}

\makeatletter

\def\instrefs#1{{\def\scsep{\def\scsep{,}}\@for\w:=#1\do{\scsep\ref{inst:\w}}}}

\renewcommand{\inst}[1]{\unskip$^{\instrefs{#1}}$}

\let\orgautoref\autoref
\renewcommand{\autoref}
        {\def\equationautorefname{Eq.}
         \def\figureautorefname{Fig.}
         \def\sectionautorefname{Sect.}
         \def\subsectionautorefname{Sect.}
         \def\subsubsectionautorefname{Sect.}
         \orgautoref}

\renewcommand*\aa@pageof{, page \thepage{} of \pageref*{LastPage}}

\makeatother

\begin{document}

\title{Precise mass and radius of a transiting super-Earth planet\\ orbiting the M dwarf TOI-1235: a planet in the radius gap?} 
\titlerunning{Super-Earth in the radius gap of an M dwarf?}

\author{
P.~Bluhm\inst{lsw}
\and R.~Luque\inst{iac,ull}  
\and N.~Espinoza\inst{stsci}
\and E.~Pall\'e\inst{iac,ull} 
\and J.~A.~Caballero\inst{cabesac}
\and S.~Dreizler\inst{iag}
\and J.\,H.~Livingston\inst{utokyo}
\and S.~Mathur\inst{iac,ull}
\and A.~Quirrenbach\inst{lsw}
\and S.~Stock\inst{lsw}
\and V.~Van~Eylen\inst{dacp}
\and G.~Nowak\inst{iac,ull}
\and E.\,D.~L\'opez\inst{nasa,gsfc}
\and Sz.~Csizmadia\inst{dlr}
\and M.\,R.~Zapatero~Osorio\inst{cabinta}
\and P.~Sch\"ofer\inst{iag}
\and J.~Lillo-Box\inst{cabesac}
\and M.~Oshagh\inst{iac,ull}
\and E.~Gonz\'alez-\'Alvarez\inst{cabinta}
\and P.\,J.~Amado\inst{iaa}
\and D.~Barrado\inst{cabesac}
\and V.\,J.\,S.~B\'ejar\inst{iac,ull}
\and B.~Cale\inst{gmu}
\and P.~Chaturvedi\inst{tls}
\and C.~Cifuentes\inst{cabesac}
\and W.\,D.~Cochran\inst{utexas} 
\and K.\,A.~Collins\inst{cfa}
\and K.\,I.~Collins\inst{gmu}
\and M.~Cort\'es-Contreras\inst{cabesac}
\and E.~D\'iez~Alonso\inst{uniovi,ictea}
\and M.~El~Mufti\inst{gmu,sudan}
\and A.~Ercolino\inst{ccao}
\and M.~Fridlund\inst{leiden,chalm}
\and E.~Gaidos\inst{hawai}
\and R.\,A.~Garc\'ia\inst{cea,aim}
\and I.~Georgieva\inst{chalm}
\and L.~Gonz\'alez-Cuesta\inst{iac,ull}
\and P.~Guerra\inst{oaa}
\and A.\,P.~Hatzes\inst{tls}
\and Th.~Henning\inst{mpia}
\and E.~Herrero\inst{ice,ieec}
\and D.~Hidalgo\inst{iac,ull}
\and G.~Isopi\inst{ccao}
\and S.\,V.~Jeffers\inst{iag}
\and J.\,M.~Jenkins\inst{Ames} 
\and E.\,L.~N.~Jensen\inst{dpa}
\and P.~K\'abath\inst{chec}
\and A.~Kaminski\inst{lsw}
\and J.~Kemmer\inst{lsw}
\and J.~Korth\inst{koln}
\and D.~Kossakowski\inst{mpia}
\and M.~K\"urster\inst{mpia}
\and M.~Lafarga\inst{ice,ieec}
\and F.~Mallia\inst{ccao}
\and D.~Montes\inst{ucm}
\and J.\,C.~Morales\inst{ice,ieec}
\and M.~Morales-Calder\'on\inst{cabesac}
\and F.~Murgas\inst{iac,ull}
\and N.~Narita\inst{iac,komaba,jst,abc}
\and V.\,M.~Passegger\inst{hs,uok}
\and S.~Pedraz\inst{caha}
\and C.\,M.~Persson\inst{chalm}
\and P.~Plavchan\inst{gmu}
\and H.~Rauer\inst{dlr,tub,fu}
\and S.~Redfield\inst{wes}
\and S.~Reffert\inst{lsw}
\and A.~Reiners\inst{iag}
\and I.~Ribas\inst{ice,ieec}
\and G.\,R.~Ricker\inst{kavli}
\and C.~Rodr\'iguez-L\'opez\inst{iaa}
\and A.\,R.~G.~Santos\inst{ssi}
\and S.~Seager\inst{kavli,MIT1,MIT2}
\and M.~Schlecker\inst{mpia}
\and A.~Schweitzer\inst{hs}
\and Y.~Shan\inst{iag}
\and M.~G.~Soto\inst{qmul}
\and J.~Subjak\inst{chec}
\and L.~Tal-Or\inst{ariel,iag}
\and T.~Trifonov\inst{mpia}
\and S.~Vanaverbeke\inst{leuven,iris}
\and R.~Vanderspek\inst{kavli}
\and J.~Wittrock\inst{gmu}
\and M.~Zechmeister\inst{iag}
\and F.~Zohrabi\inst{missi}
}

\institute{
 \label{inst:lsw}Landessternwarte, Zentrum f\"ur Astronomie der Universit\"at Heidelberg, K\"onigstuhl 12, 69117 Heidelberg, Germany
\email{pbluhm@lsw.uni-heidelberg.de}
\and 
\label{inst:iac}Instituto de Astrof\'isica de Canarias, 38205 La Laguna, Tenerife, Spain
\and 
\label{inst:ull}Departamento de Astrof\'isica, Universidad de La Laguna, 38206 La Laguna, Tenerife, Spain
\and 
\label{inst:stsci}Space Telescope Science Institute, 3700 San Martin Drive, Baltimore, MD 21218, USA
\and 
\label{inst:cabesac}Centro de Astrobiolog\'ia (CSIC-INTA), ESAC, Camino bajo del castillo s/n, 28692 Villanueva de la Ca\~nada, Madrid, Spain
\and
\label{inst:iaa}Instituto de Astrof\'isica de Andaluc\'ia (CSIC), Glorieta de la Astronom\'ia s/n, 18008 Granada, Spain
\and
\label{inst:iag}Institut f\"ur Astrophysik, Georg-August-Universit\"at, Friedrich-Hund-Platz 1, 37077 G\"ottingen, Germany
\and
\label{inst:utokyo}Department of Astronomy, University of Tokyo, 7-3-1 Hongo, Bunkyo-ky, Tokyo 113-0033, Japan
\and
\label{inst:dacp}Mullard Space Science Laboratory, University College London, Holmbury St. Mary, Dorking, Surrey, RH5 6NT, UK
\and
\label{inst:nasa}NASA Goddard Space Flight Center, 8800 Greenbelt Road, Greenbelt, MD 20771, USA
\and
\label{inst:gsfc}Sellers Exoplanet Environments Collaboration, NASA Goddard Space Flight Center, Greenbelt, MD 20771, USA
\and
\label{inst:dlr}Deutsches Zentrum f\"ur Luft- und Raumfahrt, Institut f\"ur Planetenforschung, 12489 Berlin, Rutherfordstrasse 2., Germany
\and 
\label{inst:cabinta}Centro de Astrobiolog\'ia (CSIC-INTA), Carretera de Ajalvir km 4, 28850 Torrej\'on de Ardoz, Madrid, Spain
\and
\label{inst:gmu}Department of Physics and Astronomy, George Mason University, 4400 University Drive, Fairfax, VA 22030, USA
\and 
\label{inst:tls}Th\"uringer Landessternwarte Tautenburg, Sternwarte 5, 07778 Tautenburg, Germany
\and
\label{inst:utexas}Center for Planetary Systems Habitability and McDonald Observatory, The University of Texas at Austin, Austin, TX 78730, USA
\and
\label{inst:cfa}Center for Astrophysics \textbar \ Harvard \& Smithsonian, 60 Garden Street, Cambridge, MA 02138, USA
\and 
\label{inst:uniovi}Departamento de Explotaci\'on y Prospecci\'on de Minas, Escuela de Minas, Energ\'ia y Materiales, Universidad de Oviedo, 33003 Oviedo, Spain
\and
\label{inst:ictea}Instituto Universitario de Ciencias y Tecnolog\'ias del Espacio de Asturias, Independencia 13, 33004 Oviedo, Spain
\and
\label{inst:sudan}Department of Physics, University of Khartoum, Al-Gamaa Ave, Khartoum
11111, Sudan.
\and
\label{inst:ccao}Campo Catino Astronomical Observatory, Regione Lazio, 03010 Guarcino (FR), Italy
\and
\label{inst:leiden}Leiden Observatory, Leiden University, 2333CA Leiden, The Netherlands
\and
\label{inst:chalm}Department of Space, Earth and Environment,
Chalmers University of Technology, Onsala Space Observatory, 439 92 Onsala, Sweden
\and
\label{inst:hawai}Department of Earth Sciences, University of Hawai'i at M\={a}anoa, Honolulu, HI 96822 USA
\and
\label{inst:cea}IRFU, CEA, Universit\'e Paris-Saclay, 91191 Gif-sur-Yvette, France
\and
\label{inst:aim}AIM, CEA, CNRS, Universit\'e Paris-Saclay, Universit\'e Paris Diderot, Sorbonne Paris Cit\'e, 91191 Gif-sur-Yvette, France
\and
\label{inst:oaa}Observatori Astron\`{o}mic Albany\`{a}, Cam\'i de Bassegoda s/n, 17733 Albany\`{a}, Girona, Spain
\and
\label{inst:mpia}Max-Planck-Institut f\"ur Astronomie, K\"onigstuhl 17, 69117 Heidelberg, Germany
\and
\label{inst:ice}Institut de Ci\`encies de l’Espai (ICE, CSIC), Campus UAB, Can Magrans s/n, 08193 Bellaterra, Spain
\and 
\label{inst:ieec}Institut d’Estudis Espacials de Catalunya (IEEC), 08034 Barcelona, Spain
\and
\label{inst:Ames}NASA Ames Research Center, Moffett Field, CA 94035, USA
\and
\label{inst:dpa}Department of Physics \& Astronomy, Swarthmore College, Swarthmore PA 19081, USA
\and
\label{inst:chec}Astronomical Institute, Czech Academy of Sciences, Fri\v{c}ova 298, 25165,
Ond\v{r}ejov, Czech Republic
\and
\label{inst:koln}Rheinisches Institut f\"ur Umweltforschung an der Universit\"at zu K\"oln, Aachener Strasse 209, 50931 K\"oln, Germany
\and
\label{inst:ucm}Departamento de F\'{i}sica de la Tierra y Astrof\'{i}sica and IPARCOS-UCM (Instituto de F\'{i}sica de Part\'{i}culas y del Cosmos de la UCM), Facultad de Ciencias F\'{i}sicas, Universidad Complutense de Madrid, 28040, Madrid, Spain
\and
\label{inst:komaba}Komaba Institute for Science, The University of Tokyo, 3-8-1 Komaba, Meguro, Tokyo 153-8902, Japan
\and
\label{inst:jst}JST, PRESTO, 3-8-1 Komaba, Meguro, Tokyo 153-8902, Japan
\and
\label{inst:abc}Astrobiology Center, 2-21-1 Osawa, Mitaka, Tokyo 181-8588, Japan
\and 
\label{inst:hs}Hamburger Sternwarte, Universit\"at Hamburg, Gojenbergsweg 112, 21029 Hamburg, Germany
\and
\label{inst:uok}Homer L. Dodge Department of Physics and Astronomy, University of Oklahoma, 440 West Brooks Street, Norman, OK 73019, USA
\and
\label{inst:caha}Centro Astron\'omico Hispano-Alem\'an, Observatorio de Calar Alto, Sierra de los Filabres, 04550 G\'ergal, Spain
\and
\label{inst:tub}Zentrum f\"ur Astronomie und Astrophysik, Technische Universit\"at
Berlin, Hardenbergstr. 36, 10623 Berlin, Germany
\and
\label{inst:fu}Institut f\"ur Geologische Wissenschaften, Freie Universit\"at Berlin, Malteserstr. 74–100, 12249 Berlin, Germany
\and
\label{inst:wes}Astronomy Department and Van Vleck Observatory, Wesleyan University, Middletown, CT 06459, USA
\and
\label{inst:kavli}Kavli Institute for Astrophysics and Space Research, Massachusetts Institute of Technology, Cambridge, MA 02139, USA
\and
\label{inst:ssi}Space Science Institute, 4765 Walnut St., Suite B, Boulder, CO 80301, USA
\and
\label{inst:MIT1}Department of Earth, Atmospheric and Planetary Sciences, Massachusetts Institute of Technology, Cambridge, MA 02139, USA
\and
\label{inst:MIT2}Department of Aeronautics and Astronautics, Massachusetts Institute of Technology, 77 Massachusetts Avenue, Cambridge, MA 02139, USA
\and
\label{inst:qmul}School of Physics and Astronomy, Queen Mary University London, 327 Mile End Road, London E1 4NS, UK
\and 
\label{inst:ariel}Department of Physics, Ariel University, Ariel 40700, Israel
\and
\label{inst:leuven}Vereniging Voor Sterrenkunde, Brugge, Belgium \& Centre for mathematical Plasma-Astrophysics, Department of Mathematics, KU Leuven, Celestijnenlaan 200B, 3001 Heverlee, Belgium
\and
\label{inst:iris}AstroLAB IRIS, Provinciaal Domein ``De Palingbeek'', Verbrandemolenstraat 5, 8902 Zillebeke, Ieper, Belgium
\and
\label{inst:missi}Department of Physics and Astronomy, Mississippi State University, 355 Lee Boulevard, Mississippi State, MS 39762, USA
}

\date{Received 13 April 2020 / Accepted dd Month 2020}

\abstract{We report the confirmation of a transiting planet around the bright weakly active M0.5\,V star TOI-1235 (TYC~4384--1735--1, $V \approx$ 11.5\,mag), whose transit signal was detected in the photometric time series of sectors 14, 20, and 21 of the \textit{TESS} space mission. 
We confirm the planetary nature of the transit signal, which has a period of 3.44\,d, by using precise RV measurements with the CARMENES, HARPS-N, and iSHELL spectrographs, supplemented by high-resolution imaging and ground-based photometry. A comparison of the properties derived for TOI-1235\,b with theoretical models reveals that the planet has a rocky composition, with a bulk density slightly higher than that of Earth. In particular, we measure a mass of $M_{\rm p}$~=~5.9$\pm$0.6\,$M_{\oplus}$ and a radius of $R_{\rm p}$~=~1.69$\pm$0.08\,$R_{\oplus}$, which together result in a density of $\rho_{\rm p}$~=~$6.7 ^{+ 1.3}_{- 1.1}$\,g\,cm$^{-3}$.  
When compared with other well-characterized exoplanetary systems, the particular combination of planetary radius and mass places our discovery in the radius gap, which is a transition region between rocky planets and planets with significant atmospheric envelopes. A few examples of planets occupying the radius gap are known to date. While the exact location of the radius gap for M dwarfs is still a matter of debate, our results constrain it to be located at around 1.7\,$R_\oplus$ or larger at the insolation levels received by TOI-1235\,b ($\sim$60\,$S_\oplus$). This makes it an extremely interesting object for further studies of planet formation and atmospheric evolution.
}
\keywords{planetary systems --
    techniques: photometric --
    techniques: radial velocities --
    stars: individual: TOI-1235 --
    stars: late-type
    }

\maketitle

\section{Introduction}\label{sec:intro}

Currently, over 4000 exoplanetary systems have been discovered orbiting stars other than the Sun\footnote{\url{https://exoplanetarchive.ipac.caltech.edu/}, \\\url{http://exoplanet.eu/}}, with the majority of the planets having sizes between that of the Earth and Neptune \citep{2013ApJS..204...24B}. 
Most of these systems were discovered by the \textit{Kepler} mission \citep{KEPLER,KEPLER2016}, which by design focused its transit survey on stars of spectral types F, G, and K. 
In order to understand the processes involved in the formation and evolution of planets, it is useful to compare the variations in the outcomes in different environments, for instance, by considering planetary demographics in a range of host star contexts. 
No picture of exoplanet populations can be complete without a sizable and representative sample of planetary systems around M dwarfs, which are the most common type of stars in our Galaxy \citep{2003PASP..115..763C,2006AJ....132.2360H}.

The occurrence rate of small planets orbiting M dwarfs indeed appears to increase toward late spectral subtypes at all orbital periods \citep{2013A&A...549A.109B,2015ApJ...807...45D,2015ApJ...798..112M,2016MNRAS.457.2877G}. 
In spite of this abundance, the number of exoplanets with M-star hosts whose radii and masses are precisely known is still small because these stars are intrinsically faint, and only the closest ones are well suited for detailed follow-up and characterization.

One of the most interesting features observed in the distribution of sizes of small ($R<$ 4\,$R_\oplus$) exoplanets has been the bimodal nature of this distribution, which is commonly referred to as the  ``radius gap''.
It separates planets with radii slightly smaller than that of Neptune (2--4\,$R_\oplus$) from those with radii slightly larger than Earth (1--2\,$R_\oplus$). 
While the former are believed to bear a significant contribution of water \citep{2018arXiv180306708M}, the latter are thought to be predominantly rocky. 
Although it was theoretically predicted \citep[e.g.,][]{owen2013,Jin2014,2014ApJ...792....1L,Chen2016}, the radius gap was observationally characterized only relatively recently \citep[e.g.,][]{Fulton17,Zeng2017,vaneylen2018,2018ApJ...866...99B,Fulton18} owing to an improvement in the planetary radius determination through more accurate models and stellar radii. This was possible through new high-resolution stellar spectroscopy \citep{2019A&A...625A..68S}, asteroseismology \citep{2019LRSP...16....4G}, and precise parallactic distances from the \textit{Gaia} mission \citep{GaiaDR2}. 

Two classes of models are currently accepted to explain this radius gap: photoevaporation models, which posit that planets that finally lie below the 
radius gap lost their atmospheres due to X-ray and ultraviolet radiation from the star \citep[XUV; e.g.,][]{owen2013,Lopez2013,Jin2014,Chen2016,Owen2017}, and core-powered mass-loss models, which also propose that close-in planets below the radius gap have lost their atmospheres, but conjecture that mass loss is powered by heat from the planetary core  \citep{Ginzburg+2016,Ginzburg+2018,2019MNRAS.487...24G}. These two mechanisms have different dependences on the stellar type of the host stars and the total irradiation that the planets receive \citep{wu2019,gupta2020}, which means that the actual location of the radius gap can indeed change with these parameters. 
Because most of the existing studies are based on \textit{Kepler} samples or subsamples, which are samples that are heavily focused on F, G and K-type stars, transiting exoplanetary systems around M-type stars have a huge potential to help constrain the most important mechanism(s) producing this bimodal distribution \citep[see, e.g.,][]{Hirano2018}. 
Measuring the planetary mass in turn allows us to gain some insight into the bulk composition of the exoplanets, which delivers a clearer picture of the underlying nature of the radius gap. 
The {\it Transiting Exoplanet Survey Satellite} 
\citep[{\it TESS};][]{Ricker2015} has proven to be a prime instrument for detecting and characterizing small planets orbiting bright stellar hosts. 
Having completed its first year of monitoring, it has contributed to the detection and confirmation of more than 40 new transiting exoplanetary systems,  many of which consist of small planets orbiting low-mass M stars \citep[e.g.,][]{2019A&A...628A..39L, 2019ApJ...883L..16C, 2019NatAs...3.1099G, 2020A&A...636A..58A, 
2020AJ....160....3C, 2020arXiv200100952G, 2020arXiv200301140N}.
Here we report on a very interesting addition to this growing sample of {\em TESS} transiting exoplanet discoveries around M dwarfs: 
a transiting super-Earth that appears to be right in the radius gap for low-mass stars orbiting the early M dwarf \object{TOI-1235} (see also the coordinated, but intentionally independent, announcement by \citealt{2020AJ....160...22C}). 

The paper is organized as follows. Section \ref{sec:data} presents the {\it TESS} photometry we used, along with ground-based observations of the star, including high-resolution spectroscopy, lucky and speckle imaging, and photometric variability monitoring. 
Section~\ref{sec:star} presents the stellar properties of the host star, newly derived and collected from the literature. 
In Section~\ref{sec:analysis} we present our analysis of the available data to constrain the planetary properties of the system.
In Section~\ref{sec:discussion} we discuss our results, with an emphasis on the location of the planet in the mass-radius diagram and its composition, and, finally, Section~\ref{sec:conclusions} shows our conclusions.

\section{Data} \label{sec:data}

Radius and mass are key physical properties of a planet. 
Together, they inform the planetary density, bulk composition, internal structure, and ability to retain an atmosphere. 
The combination of a transit and radial velocity (RV) detection is the most straightforward way to measure both the planetary radius and mass.
In this work, we used the CARMENES\footnote{Calar Alto high-Resolution search for M dwarfs with Exoearths with Near-infrared and optical Echelle Spectrographs: \url{http://carmenes.caha.es}}, HARPS-N\footnote{ High Accuracy Radial velocity Planet Searcher for the Northern hemisphere: \url{https://plone.unige.ch/HARPS-N/}}, and iSHELL\footnote{Immersion Grating Echelle Spectrograph: \url{http://irtfweb.ifa.hawaii.edu/~ishell/}} high-resolution spectrographs for the RV follow-up (Sect.~\ref{subsec:rv}).

Moreover, given the intrinsic faintness of M dwarfs in general and the large photometric apertures of wide-field surveys such as {\em TESS} in particular, many light curves with transit candidates are susceptible to contamination by nearby sources.
Blends with stars other than the target star are frequent, especially at low Galactic latitudes, while many stars are unresolved multiples.
In some cases, other stars in the aperture mask are variable and bright enough to affect the photometric, and even RV, measurements.
A particularly difficult type of false positives are background-eclipsing binaries near the target star, which can mimic planet transits.
High-resolution imaging follow-up is therefore needed to identify nearby potential contaminants, and ground-based photometric monitoring is helpful in discarding false positives, such as nearby eclipsing binaries.
For this second follow-up stage, we used the lucky and speckle imagers AstraLux and NESSI and collected light curves complementary to those of {\em TESS}, either measured by us with 1\,m class telescopes (LCOGT, TJO) or compiled from public data bases (Sects.~\ref{subsec:hiresim} and~\ref{subsec:phot}).

\begin{table}
\centering
\small
\caption{{\em TESS} observations of TOI-1235.} 
\label{tab:phot}
\begin{tabular}{cccll}
\hline\hline
\noalign{\smallskip}
Sector   &  Camera   & CCD   & Start date & End date\\
\noalign{\smallskip}
\hline
\noalign{\smallskip}
14  &  4  &   3  &  18 July 2019 &  15 August 2019\\
20  &  2  &   1  &  24 December 2019 & 21 January 2020\\
21  &  2  &   2  &  21 January 2020 & 18 February 2020\\
\noalign{\smallskip}         
\hline
\end{tabular}
\end{table}

\subsection{TESS photometry} \label{subsec:transit}

\begin{figure*}
    \centering
    \includegraphics[width=\textwidth]{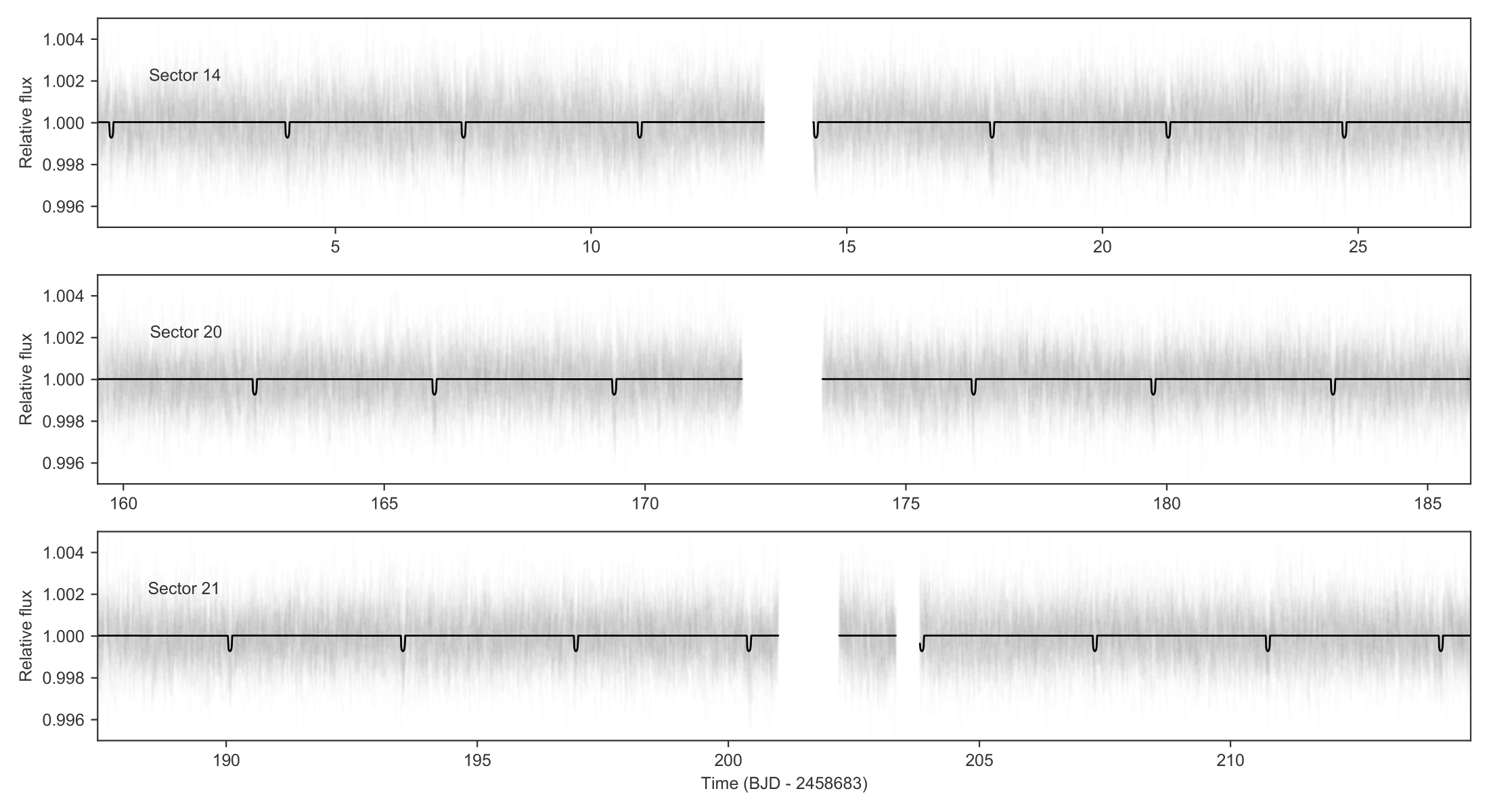}
    \includegraphics[width=0.8\textwidth]{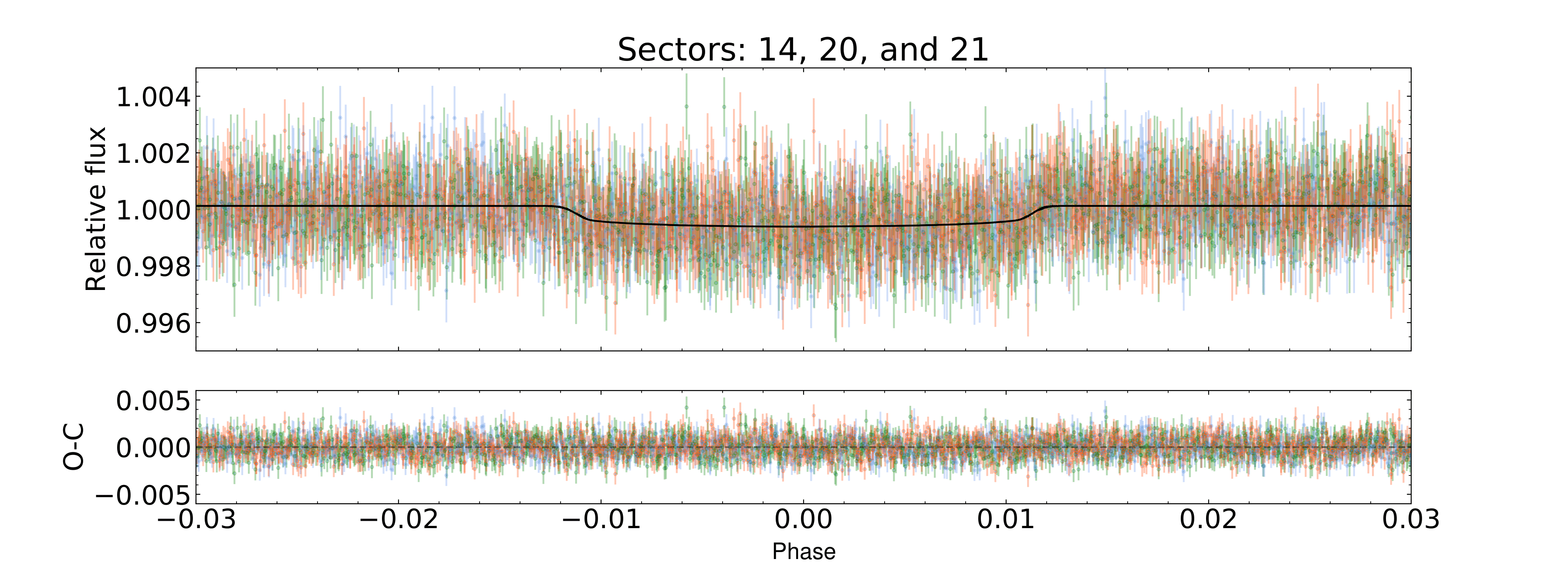}    
    \caption{{\em Top panel}: \textit{TESS} transit photometry for the three sectors (gray points) with the best-fit \texttt{juliet} model (black line; see Sect.~\ref{subsec:joint} for details on the modeling). 
    {\em Bottom panel}: Phase-folded transit light curve of TOI-1235 b. 
    The three sectors (14, 20, and 21) are color-coded in red, green, and blue, respectively.
    The best-fit model is depicted by the black curve.}
    \label{fig:photo-TESS}
\end{figure*}

The goal of \textit{TESS} is to search for planets transiting bright and nearby stars. 
It was designed to observe 26 sectors of 24\,$\times$\,96\,deg$^2$ split into 13 partially overlapping sectors in the 
north and south ecliptic hemispheres, each of which observed for 27--28\,d.
The Mikulski Archive for Space Telescopes\footnote{\url{https://mast.stsci.edu}, \url{https://archive.stsci.edu/}} (MAST) stores the light curves of stars in the \textit{TESS} Input Catalog \citep[TIC;][]{2018AJ....156..102S}.
About 200,000 bright main-sequence F, G, K, and M-type stars, small enough to enable the detection of transiting planets smaller than Neptune ($R \leq 4\,R_{\oplus}$), are observed with a 2\,min cadence \citep[cf.][]{2018AJ....156..102S}, while fainter stars ($V \gtrsim$ 13\,mag) or with earlier spectral types (and, thus, high masses and large radii) are observed with a 30\,min cadence.
TOI-1235 (TIC~103633434) was observed by \textit{TESS} in 2\,min short-cadence integrations in sectors 14, 20, and 21 during the \textit{TESS} primary mission (see Table~\ref{tab:phot}). 
The transiting-planet signature was detected when the Science Processing Operations Center \citep[SPOC;][]{SPOC} processed the data from sector 14 and searched the light curve for transits with the  transiting planet search module \citep{2002ApJ...575..493J, 2017ksci.rept....9J}. The transit signature passed all the diagnostic tests performed by the data validation module \citep{Twicken:DVdiagnostics2018}, which also provided the initial limb-darkened transit model fit \citep{Li:DVmodelFit2019}.
After these steps, TOI-1235 was announced on 16 September 2019 as a \textit{TESS} object of interest (TOI) thorugh the dedicated MIT \textit{TESS} data alerts public website\footnote{\url{https://tess.mit.edu/toi-releases/}}.
The transit signal had a period of 3.4431$\pm$0.0008\,d and a depth of 0.91$\pm$0.08\,mmag, corresponding to a planet radius of about 2\,$R_{\oplus}$, well in the super-Earth domain.

Soon after, we downloaded the corresponding light curve produced by the SPOC at the NASA Ames Research Center from MAST.
SPOC provides simple aperture photometry (SAP) and photometry corrected for systematics \citep[PDC,][]{2012PASP..124.1000S,Stumpe2012PASP..124..985S,Stumpe2014PASP..126..100S}, which is optimized for \textit{TESS} transit searches. Figure~\ref{fig:photo-TESS} shows the PDC data for the three \textit{TESS} sectors with the best-fit model (see Sect.~\ref{subsec:joint} for details).

\subsection{High-resolution spectroscopy}\label{subsec:rv}

\subsubsection{CARMENES}\label{subsec:carmenes}

CARMENES \citep{CARMENES, CARMENES18} is a high-resolution spectrograph mounted on the 3.5\,m telescope at the Observatorio de Calar Alto in Almer\'ia, Spain. 
It splits the incoming light into two channels, one that operates in the optical (VIS:~$0.52$--$\SI{0.96}{\micro\metre}$, $\mathcal{R}=94\,600$) and the other in the near-infrared (NIR:~$0.96$--$\SI{1.71}{\micro\metre}$, $\mathcal{R}=80\,400$). 
TOI-1235 was observed 40 times with CARMENES between 09~November~2019 and 18~February~2020, overlapping with the {\em TESS} sector 20 and 21 observations.
We used the maximum exposure time of 1800\,s and followed the standard data flow of the CARMENES guaranteed time observations.
In particular, we reduced the VIS spectra with \texttt{CARACAL} \citep{2014A&A...561A..59Z} and determined the corresponding RVs and spectral activity indices (see Sect.~\ref{subsec:GLS_rv}) with \texttt{SERVAL} \citep{2018A&A...609A..12Z}.
We corrected the RV’s for barycentric motion, instrumental drift, secular acceleration, and nightly zero-points (see \citealt{Kaminski18}, \citealt{2019MNRAS.484L...8T},
and, especially, \citealt{2020A&A...636A..74T} for details).
For exposure times shorter than 1800\,s, the CARMENES standard integrations are automatically limited by signal-to-noise (S/N) ratio to 150 by an exposure-meter per channel that collects the light of the zeroth order of the respective \'echelle grating during the exposure \citep{2016SPIE.9910E..0EC}.
However, the median S/N of our CARMENES VIS spectra was slightly lower ($\sim$97).
Correspondingly, the weighted root-mean-square (wrms) and median uncertainty ($\hat{\sigma}$) of the CARMENES VIS data were 3.7\,m\,s$^{-1}$ and 1.9\,m\,s$^{-1}$, respectively.
These RVs and their uncertainties are listed in the top part of Table~\ref{tab:RV_Activity_all}.

As expected from the results presented by \citet{2020arXiv200601684B}, the RV precision of the CARMENES NIR observations of TOI-1235 was lower than that of the VIS observations: wrms = 9.1\,m\,s$^{-1}$ and $\hat{\sigma}$ = 7.4\,m\,s$^{-1}$.
The expected RV amplitude of the planet of about 3--4\,m\,s$^{-1}$ was lower than the data radial precision of the CARMENES NIR data.
The RVs, CRX, and dLW of CARMENES NIR spectra are displayed in the top part of Table~\ref{tab:RV_NIR}.

\subsubsection{HARPS-N}\label{subsec:harpsN}

HARPS-N \citep{2012SPIE.8446E..1VC} is a high-resolution spectrograph mounted on the Italian 3.58\,m Telescopio Nazionale Galileo at the Observatorio del Roque de los Muchachos, La Palma, Spain.
HARPS-N covers the optical wavelength regime between 0.38\,$\mu$m and 0.69\,$\mu$m with a spectral resolution of $\mathcal{R}=115\,000$. 
The precision and stability of HARPS-N is comparable to its sister instrument HARPS on the ESO~3.6\,m telescope and therefore to CARMENES \citep{2018A&A...609A.117T, 2019A&A...624A.123P}.
TOI-1235 was observed 21 times between 14 January 2020 and 26 February 2020 with HARPS-N\footnote{HARPS-N data: 15 RVs were obtained from the Spanish CAT19A-162 program (PI: Nowak) and 6 RVs from ITP\,19-1 program (PI: Pall\'e).}, also overlapping with {\em TESS} sectors 20 and 21.
Just as with the CARMENES data, we determined the RVs and H$\alpha$ spectral activity index with \texttt{SERVAL}. They are listed in the bottom part of Table~\ref{tab:RV_Activity_all}.
The typical S/N per exposure was 100, while the wrms and $\hat{\sigma}$ of the HARPS-N data were 4.5\,m\,s$^{-1}$ and 1.0\,m\,s$^{-1}$, respectively.

\subsubsection{iSHELL}\label{subsec:iSHELL}

We obtained 49 spectra during five nights for TOI 1235 spanning 26 days in January-February 2020 with iSHELL mounted on the 3.2\,m NASA Infrared Telescope Facility (IRTF) on Maunakea, Hawaii \citep{2016SPIE.9908E..84R}.  
We used the silicon immersion grating optimized for the $K$ band with the 0.375\,arcsec slit, which resulted in a spectral resolution of 80\,000. The spectra were wavelength calibrated with a methane isotopolog gas cell in the calibration unit. The exposure times were 300\,s, repeated 9--11 times within a night to reach a cumulative photon S/N per spectral pixel at about $2.4\,\mu$m (at the approximate center of the blaze for the middle order) varying from 77 to 98 to achieve a per-night precision of 4--11\,m\,s$^{-1}$. 
Spectra were reduced and RVs extracted using the methods outlined by \citet{2019AJ....158..170C}. 
The resulting wrms and $\hat{\sigma}$ of the iSHELL data were 7.2\,m\,s$^{-1}$ and 6.1\,m\,s$^{-1}$, slightly better than the CARMENES NIR data, but still twice higher than the expected planet semiamplitude. 
The RVs are displayed in the bottom part of Table~\ref{tab:RV_NIR}.

\subsection{High-resolution imaging} \label{subsec:hiresim}

\subsubsection{AstraLux} \label{subsec:lucky}

We observed TOI-1235 with the high spatial resolution camera and lucky imager AstraLux \citep{hormuth08} on the 2.2\,m telescope at the Observatorio de Calar Alto in Almer\'ia, Spain.
The observations were carried out in the $z'$ band on 30 October 2019 under good weather conditions with a mean seeing of 1.0\,arcsec. 
We obtained 96\,000 frames of 10\,ms in a $6.0 \times 6.0$\,arcsec$^2$ window.
With the observatory pipeline, we selected the 5\,\% frames with the highest Strehl ratio \citep{strehl1902}, aligned them, and stacked them for a final high-spatial resolution image. 

\subsubsection{NESSI} \label{subsec:speckle}

On 14 October of 2019, we observed TOI-1235 with the NASA Exoplanet Star and Speckle Imager \citep[NESSI;][]{2018PASP..130e4502S, 2018SPIE10701E..0GS} on the 3.5\,m WIYN telescope at the Kitt Peak National Observatory in Arizona, USA. 
We observed nearby point-source calibrator stars and reduced the data following \citet{2011AJ....142...19H}. The high-speed electron-multiplying CCDs of NESSI capture images at 25\,Hz simultaneously in two bands centered at 562\,nm and 832\,nm. 
Finally, we obtained two $4.6\times4.6$\,arcsec$^2$ reconstructed images, one for each passband.

\subsection{Ground-based photometry}\label{subsec:phot}

\begin{table*}
    \centering
    \small
    \caption{Descriptions of data from public ground-based surveys$^a$.} \label{tab:ground.phot}
    \begin{tabular}{llllcccccc}
        \hline\hline
        \noalign{\smallskip}
Survey & Band & Start date & End date & $N$ & $\Delta t$ & $\overline{m}$ & $\sigma_m$ & $\overline{\delta {m}}$ \\
 & & & & & (d) & (mag) & (mag) & (mag) \\ 
        \noalign{\smallskip}
        \hline
        \noalign{\smallskip}
    ASAS-SN & $g'$ & 29 October 2017 & 24 March 2020 & 603$^b$ & 877 & 12.255 & 0.026 & 0.010 \\
     & $V$   & 28 January 2012 & 26 November 2018 & 713$^b$ & 2494 & 11.572 & 0.018 & 0.009 \\ 
    NSVS & Clear & 04 June 2018 & 20 May 2019 & 111 & 359 & 11.027 & 0.024 & 0.011 \\
    Catalina$^c$ & Clear & 02 February 2006 & 18 April 2013 & 43 & 2632 & 10.761 & 0.089 & 0.050 \\
    \noalign{\smallskip}         
        \hline
    \end{tabular}
   \tablefoot{
   \tablefoottext{a}{Number of collected data points.}
   \tablefoottext{b}{After discarding 20 $g'$ and 10 $V$ dubious data points (with poor quality flags).}
   \tablefoottext{c}{Data set eventually not used.}
   }
\end{table*}

Additional photometric data for TOI-1235 were taken on 31 December 2019 with one of the 1\,m telescopes of the Las Cumbres Observatory Global Telescope \citep[LCOGT;][]{Brown:2013} Network at the McDonald Observatory in Texas, USA. 
We used the {\tt TESS Transit Finder}, which is a customized version of the {\tt Tapir} software package \citep{Jensen:2013}, to schedule a full transit observation.
We used the $zs$ (short $z'$) band and an aperture radius of 7.0\,arcsec for the photometry extraction. 
A total of 358 photometric measurements were obtained with a cadence of 56\,s and a median precision of 1.1\,mmag per point. 
The images were calibrated using the standard LCOGT {\tt Banzai}  pipeline \citep{McCully:2018}, and the photometric data were extracted using the {\tt AstroImageJ} software package \citep{Collins:2017}.

We also observed a TOI-1235 transit on 29 March 2020 with the 0.8\,m Telescopi Joan Or\'o (TJO) at the Observatori Astron\`omic del Montsec in Lleida, Spain. 
We obtained a total of 221 images with the Johnson $R$ filter using the LAIA imager, a 4k\,$\times$\,4k CCD with a field of view of 30\,arcmin and a scale of 0.4\,arcsec\,pixel$^{-1}$. 
The observations were affected by poor weather conditions, and the photometry was extracted and analyzed with {\tt AstroImageJ}. Although we did not use this photometry in the joint modeling due to the relatively poor photometric precision, of about 2\,mmag, and an observational gap in the middle of the transit, it was still useful as an independent confirmation that the transit event indeed occurred on the target star, as the TJO photometry for all {\em Gaia} DR2 sources within 2.5\,arcmin of the target ruled out the possibility that the {\em TESS} transit signal was produced by any of these stars being short-period eclipsing binary contaminants. 

Finally, we searched for public time-series data of wide-area photometric surveys and databases exactly as in \citet{2019A&A...621A.126D}.
In particular, we retrieved light curves from the All-Sky Automated Survey for SuperNovae \citep[ASAS-SN;][]{2014ApJ...788...48S,2017PASP..129j4502K} in the $g'$ and $V$ bands, and the Northern Sky Variability Survey \citep[NSVS;][]{2004AJ....127.2436W}, and the Catalina Sky Survey \citep{2009ApJ...696..870D} in white light.
Table~\ref{tab:ground.phot} summarizes the three public data sets, including the standard deviation of the magnitudes and mean of the magnitude errors.
The Catalina data set is much noisier, sparser, and shorter than the others, therefore we did not use it in our analysis.
In addition, we did not find data on TOI-1235 in the public archives of other photometric surveys, such as
MEarth \citep{2011ApJ...727...56I},
SuperWASP \citep[][including unpublished data]{2006PASP..118.1407P},
ASAS \citep{1997AcA....47..467P},
and HATNet \citep{2004PASP..116..266B}.
Finally, TOI-1235 was not labeled as a variable star in the ATLAS survey \citep{2018AJ....156..241H}.

\section{Stellar properties} \label{sec:star}

\begin{table}
\centering
\small
\caption{Stellar parameters of TOI-1235.} \label{tab:star}
\begin{tabular}{lcr}
\hline\hline
\noalign{\smallskip}
Parameter   & Value             & Reference \\ 
\hline
\noalign{\smallskip}
\multicolumn{3}{c}{\em Name and identifiers}\\
\noalign{\smallskip}
Name    & TYC~4384--1735--1     & H{\o}g00 \\
Karmn   & J10088+692            & AF15 \\
TOI     & 1235                  & ExoFOP-TESS \\  
TIC     & 103633434             & Sta18 \\  
\noalign{\smallskip}
\multicolumn{3}{c}{\em Coordinates and spectral type}\\
\noalign{\smallskip}
$\alpha$ (J2000) & 10:08:52.38  & {\it Gaia} DR2 \\
$\delta$ (J2000) & +69:16:35.8  & {\it Gaia} DR2 \\
Sp. type         & M0.5\,V      & Lep13 \\
$G$ [mag]        & $10.8492\pm0.0005$ & {\it Gaia} DR2 \\
$T$ [mag]       & $9.9192\pm0.0072$ & Sta19\\
$J$ [mag]        & $8.711\pm0.020$ & Skr06 \\
\noalign{\smallskip}
\multicolumn{3}{c}{\em Parallax and kinematics}\\
\noalign{\smallskip}
$\varpi$ [mas]  & $25.202\pm0.030$ & {\it Gaia} DR2 \\
$d$ [pc]        & $39.680\pm0.048$ & {\it Gaia} DR2 \\
$\mu_{\alpha}\cos\delta$ [$\mathrm{mas\,a^{-1}}$]  & $+196.631 \pm 0.040$ & {\it Gaia} DR2 \\
$\mu_{\delta}$ [$\mathrm{mas\,a^{-1}}$] & $+17.369 \pm 0.047$ & {\it Gaia} DR2 \\
$\gamma$ [$\mathrm{km\,s^{-1}}$] & $-27.512\pm0.018$ & This work \\
$U$ [$\mathrm{km\,s^{-1}}$]  & $+45.98\pm0.04$ & This work \\
$V$ [$\mathrm{km\,s^{-1}}$]  & $-4.29\pm0.01$ & This work \\
$W$ [$\mathrm{km\,s^{-1}}$]  & $+1.73\pm0.03$ & This work \\
Gal. population & Thin disk & This work \\
\noalign{\smallskip}
\multicolumn{3}{c}{\em Photospheric parameters}\\
\noalign{\smallskip}
$T_{\mathrm{eff}}$ [K]  & $3997 \pm 51$ & This work   \\
$\log g$                & $4.64 \pm 0.04$ & This work   \\
{[Fe/H]}                & $+0.33 \pm 0.16$ & This work   \\
$v \sin i_\star$ [$\mathrm{km\,s^{-1}}$]    & $<2.0$ & This work \\
\noalign{\smallskip}
\multicolumn{3}{c}{\em Physical parameters}\\
\noalign{\smallskip}
$L_\star$ [$10^{-4}\,L_\odot$] & $883 \pm 3$      & This work \\
$M_\star$ [$M_{\odot}$]       & $0.630 \pm 0.024$ & This work \\
$R_\star$ [$R_{\odot}$]       & $0.619 \pm 0.019$ & This work \\
\noalign{\smallskip}
\multicolumn{3}{c}{\em Activity and age}\\
\noalign{\smallskip}
pEW(H$\alpha$) [\AA]    & $+0.97 \pm 0.06$ & This work \\
$\log R'_{\rm HK}$      & $-4.728 \pm 0.015$ & This work \\
$S_{\rm MWO}$           & $1.005 \pm 0.029$ & This work \\
Age (Ga)            & 0.6--10           & This work \\
\noalign{\smallskip}
\hline
\end{tabular}
\tablebib{
    AF15: \citet{2015A&A...577A.128A};
    {\it Gaia} DR2: \citet{GaiaDR2};
    H{\o}g00: \citet{2000A&A...355L..27H};
    Lep13: \citet{2013AJ....145..102L};
    Schf19: \citet{Schf19};
    Skr06: \citet{2MASS};
    Sta18: \citet{2018AJ....156..102S};
    Sta19: \citet{2019AJ....158..138S}.
}
\end{table}

The star TOI-1235 (TYC~4384--1735--1) has been included in only a few proper-motion surveys \citep{2000A&A...355L..27H, 2005AJ....129.1483L, 2016ApJS..224...36K} and catalogs of nearby M dwarfs that could host exoplanets \citep{2011AJ....142..138L, 2013MNRAS.435.2161F, 2014MNRAS.443.2561G}.
As indicated by its Tycho-2 identifier, TOI-1235 is a relatively nearby ($d \approx$ 39.6\,pc) bright ($V \approx$ 11.5\,mag) star.
\citet{2013AJ....145..102L} and \citet{2014MNRAS.443.2561G} reported spectral types M0.5\,V and M1.0\,V and effective temperatures $T_{\rm eff}$ of 3660\,K and 4060\,K.
\citet{2014MNRAS.443.2561G} also derived stellar radius $R_\star$ and bolometric luminosity $L_\star$, which are consistent with the determinations by \citet{GaiaDR2}, mass $M_\star$, and pseudo-equivalent width of the H$\alpha$ line, pEW(H$\alpha$).

We redetermined all stellar parameters for this early-M dwarf. 
In particular, we measured $T_{\rm eff}$, surface gravity $\log{g}$, and iron abundance [Fe/H] from the stacked CARMENES VIS spectra by fitting them with a grid of PHOENIX-SESAM models as in \cite{2019A&A...627A.161P},
the rotational velocity $v \sin{i}$ with the cross-correlation method as in \cite{2018A&A...612A..49R},
and the stellar luminosity $L_\star$ by integrating the spectral energy distribution as in \citet{Cifuentes.et.al.2020}.
To do this, we used photometric data in 17 passbands from the optical blue Tycho-2 $B_T$ \citep{2000A&A...355L..27H} to the mid-infrared AllWISE $W4$ \citep{2014yCat.2328....0C}, the Virtual Observatory Spectral energy distribution Analyzer \citep[VOSA;][]{2008A&A...492..277B}, and the BT-Settl CIFIST theoretical models, which were used to extrapolate the spectral energy distribution at ranges bluer than $B_T$ and redder than $W4$.
The full photometric data set including $u'$ is made available by \citet{Cifuentes.et.al.2020}.
The photospheric contributions to the total stellar flux of an M0.5\,V star in these ranges are $<$0.5\,\% and $<$\,0.004\,\%, which means that the $L_\star$ determination was model independent at the $>$99.5\,\% level at fixed metallicity. 
Next, we determined $R_\star$ through the Stefan–Boltzmann law, $L_\star = 4 \pi R{_\star}^2 \sigma T_{\rm eff}^4$, and $M_\star$ with the mass-radius relation derived from main-sequence eclipsing binaries by \citet{2019A&A...625A..68S}.
All redetermined parameters ($T_{\rm eff}$, $L{_\star}$, $R{_\star}$, and $M{_\star}$) match the values published by \citet{2014MNRAS.443.2561G} and \citet{GaiaDR2}  within $1\sigma.$ 
Furthermore, we used the precise astrometric data of {\em Gaia} DR2, the absolute RV measured on the stacked CARMENES spectra as in \citet{2020arXiv200307471L}, and the prescription of \citet{1987AJ.....93..864J} for measuring the Galactocentric space velocities $UVW$.
Using this kinematic information with the BANYAN~$\Sigma$ tool \citep{2018ApJ...856...23G}, we classified TOI-1235 as a field star in the Galactic thin disk not associated with any young stellar kinematic group.

Finally, we determined key indicators of stellar activity. 
First, we measured the Mount Wilson $S$ index, $S_{\rm MWO}$, with the {\tt Yabi} data environment on the HARPS-N spectra \citep{yabi, 2015A&A...578A..64B}, from which we derived $\log{R'_{\rm HK}}$ using the formulae of \citet{2017A&A...600A..13A} and $V-K_s$ = 3.602$\pm$0.0.059\,mag.
Next, we measured pEW(H$\alpha$) on the
CARMENES stacked spectrum following \citet{Schf19}, which was in agreement within $2\sigma$ to the pEW(H$\alpha$) = +0.74$\pm$0.11\,{\AA} measured by \citet{2014MNRAS.443.2561G} in April 2009. 
This means that the activity level of the star (as determined by H$\alpha$) has not substantially changed for over a decade.
These three indicators make TOI-1235 one of the least active stars for its spectral type \citep{2004ApJS..152..261W, 2017A&A...600A..13A, 2018A&A...616A.108B}.   
See Sect.~\ref{subsec:GLS_rv} for a search for periodic signals in other spectroscopic activity indicators.

We also searched for soft X-ray and ultraviolet data of TOI-1235, but the star was not covered by any pointing ({\em XMM-Newton}, {\em Chandra}, or {\em EUVE}), or was too faint and far from axis to be detected ({\em ROSAT} and {\em GALEX}).
As an inactive member of the thin disk without further clear evidence to support a very young or very old age, TOI-1235 is likely between 0.6\,Ga (older than the Hyades) and 10\,Ga (younger than low-metallicity thick-disk stars).

Table~\ref{tab:star} summarizes the stellar properties of TOI-1235.
We provide the average values, their uncertainties, and corresponding reference.

\section{Analysis and results}\label{sec:analysis}

\subsection{Limits on photometric contamination} \label{subsec:contamination}

\begin{figure}
    \centering
    \includegraphics[width=0.24\textwidth]{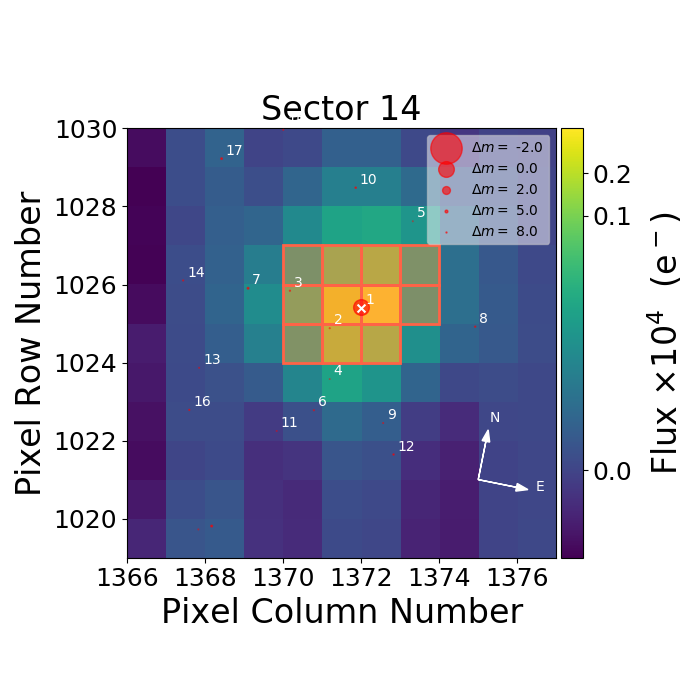}
    \includegraphics[width=0.24\textwidth]{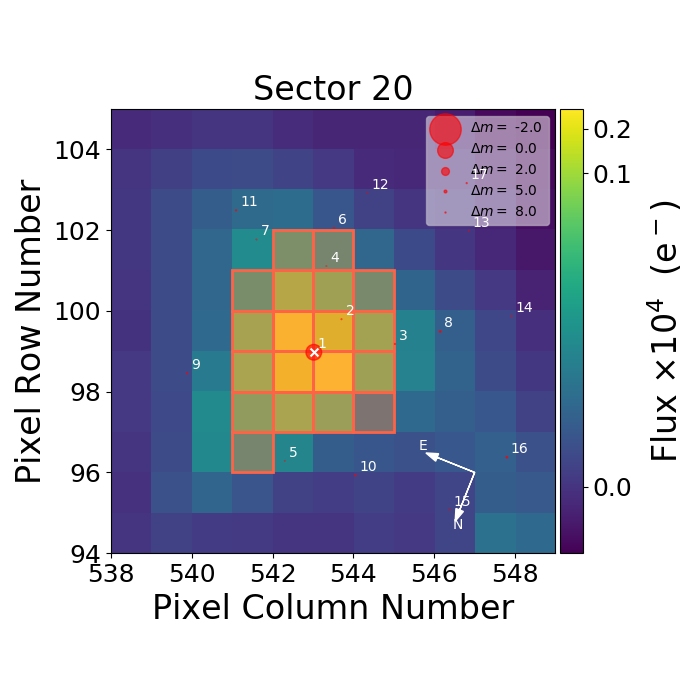}    \includegraphics[width=0.24\textwidth]{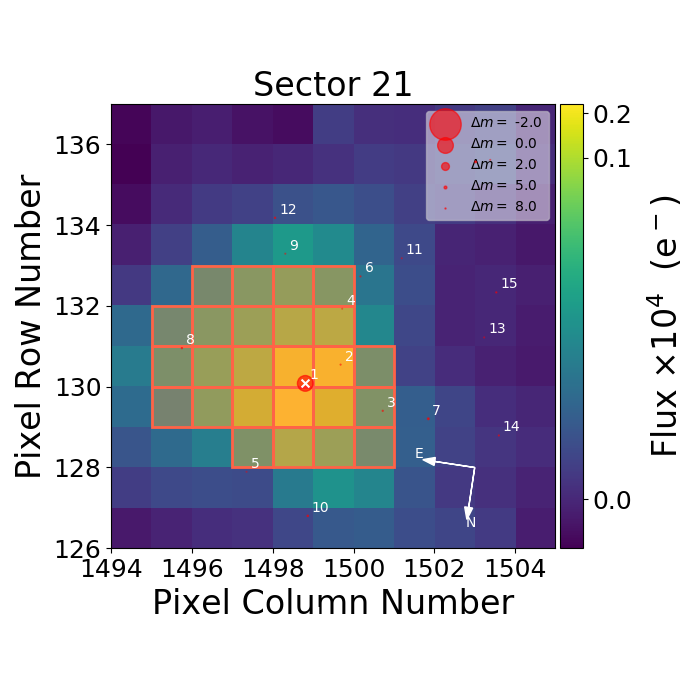} 
    \includegraphics[width=0.24\textwidth,height=0.257\textwidth]{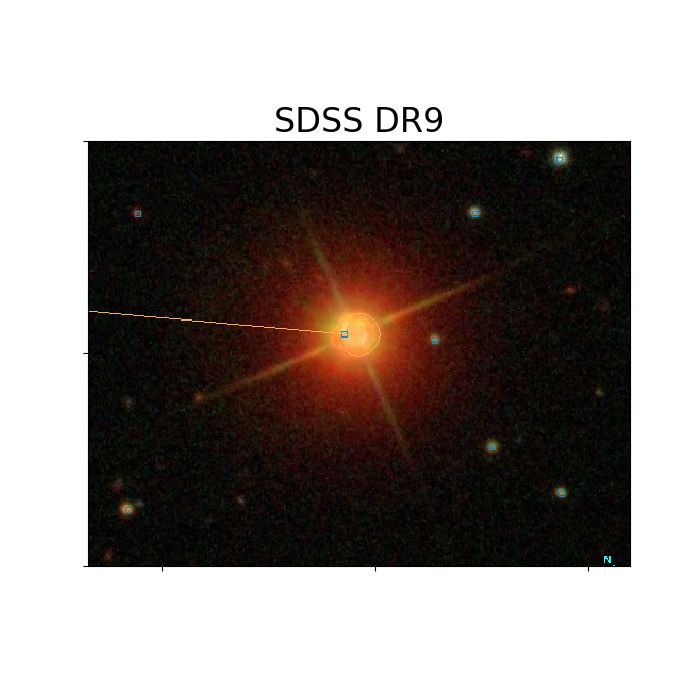}
         \caption{Target pixel files of TOI-1235 in {\em TESS} sectors 14, 20, and 21. The electron counts are color-coded. The red bordered pixels are used in the SAP. The size of the red circles indicates the {\em TESS} magnitudes of all nearby stars and TOI-1235 (circle $\text{}$1 is marked with $\text{a cross}$). 
        {\em Bottom right}: False-color, $2 \times 2$\,arcmin$^2$ Sloan Digital Sky Survey DR9 image centered on TOI-1235 (north is up, east is left).}
        \label{fig:SDSS+TPF}
\end{figure}

\begin{figure*}
    \centering
    \includegraphics[width=0.49\textwidth]{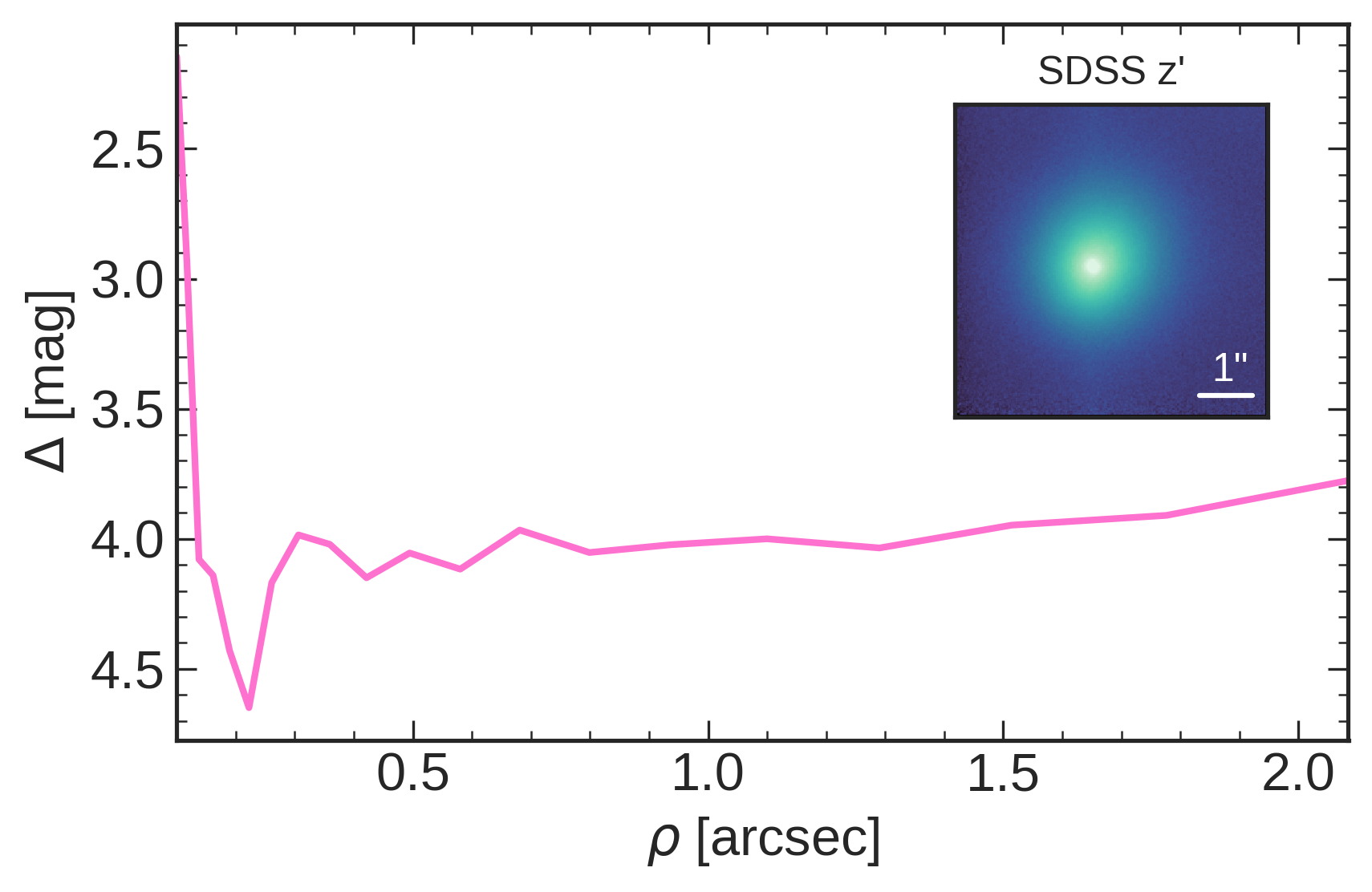}
    \includegraphics[width=0.49\textwidth]{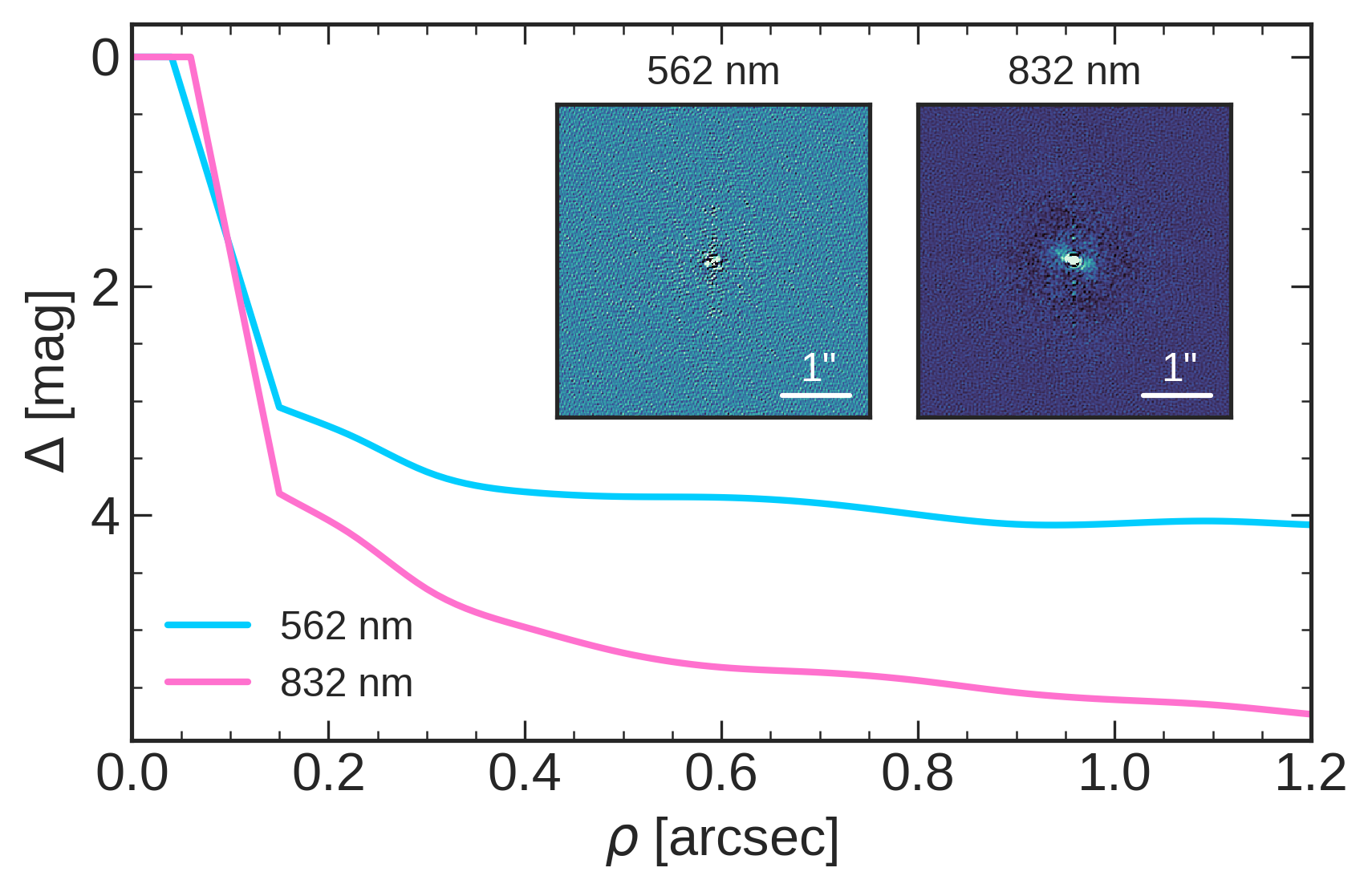}
    \caption{Contrast curves ($5 \sigma$) of TOI-1235 from AstraLux ({\em left}) and NESSI ({\em right}) observations.
        Inset images are 6.0\,$\times$\,6.0\,arcsec$^2$ stacked in $z'$ band and 4.6\,$\times$\,4.6\,arcsec$^2$ reconstructed at 562\,nm and 832\,nm, respectively.}
    \label{fig:astralux+nessi}
\end{figure*}

The presence of an unresolved companion could have a strong effect on our results. This is particularly relevant for {\em TESS} photometry because of its large pixel size ($\sim$21\,arcsec). For comparison, the CARMENES and HARPS-N optical fiber apertures projected on the sky have sizes of only 1.5\,arcsec and 1.0\,arcsec, respectively. Even so, they are not immune to contamination from sub-arcsecond blends. Here, we place limits on the dilution factor and the presence of contaminant sources that can affect our photometric and RV measurements of TOI-1235.
First, we verified that the sources in the selection apertures in the \textit{TESS} pixel file (TPF) did not affect the depth of the transits significantly. The TPFs shown in Fig.~\ref{fig:SDSS+TPF} were created with \texttt{tpfplotter}\footnote{\url{https://github.com/jlillo/tpfplotter}} \citep{2020A&A...635A.128A}. 
In particular, {\em Gaia} DR2 sources 2 and 3 in sectors 14, source 4 in sector 20, and source 8 in sector 21 all have $G$-band fluxes lower by 0.5\,\% than that of TOI-1235 (\textit{Gaia} and the \textit{TESS} photometric bands are very similar).
Similar results were found for the apertures of the ground-based surveys ASAS-SN and NSVS.

For subarcsecond separations, we used our lucky imaging AstraLux and speckle NESSI data sets described in Sect.~\ref{subsec:hiresim} and illustrated by Fig.~\ref{fig:astralux+nessi}.
We computed $5\sigma$ contrast curves as described by \cite{lillo-box12} with the \texttt{astrasens} package\footnote{\url{https://github.com/jlillo/astrasens}} for AstraLux, and as reported by \citet{2018AJ....156...78L} for NESSI.
From both data sets, we confirmed the absence of any close companion 4--6\,mag fainter than TOI-1235, and derived an upper limit to the contamination of around 2\,\% between 0.15\,arcsec and 1.5\,arcsec (6.0--60\,au if physically bound).

A further constraint came from the {\em Gaia} DR2 renormalized unit weight error ({\tt RUWE}) value, which for TOI-1235 is 1.03, below the critical value of 1.40 that ``indicates that a source is non-single or otherwise problematic for the astrometric solution'' \citep{2018A&A...616A..17A,2018A&A...616A...2L}.
We also searched for wide common proper motion companions with similar {\em Gaia} DR2 parallax, as in \citet{2018MNRAS.479.1332M}, and found none within 30\,arcmin of our star.
Following these results, we conclude that TOI-1235 is a single star. We estimated the {\em TESS} and LCOGT dilution factors at $D=1.0$ with Eq.~2 in \citet{juliet}, and fixed this value for all our model fits in the next sections.

\subsection{Stellar rotational period from photometric data} \label{subsubsec:Prot}

\begin{figure}
    \centering
    \includegraphics[width=\hsize]{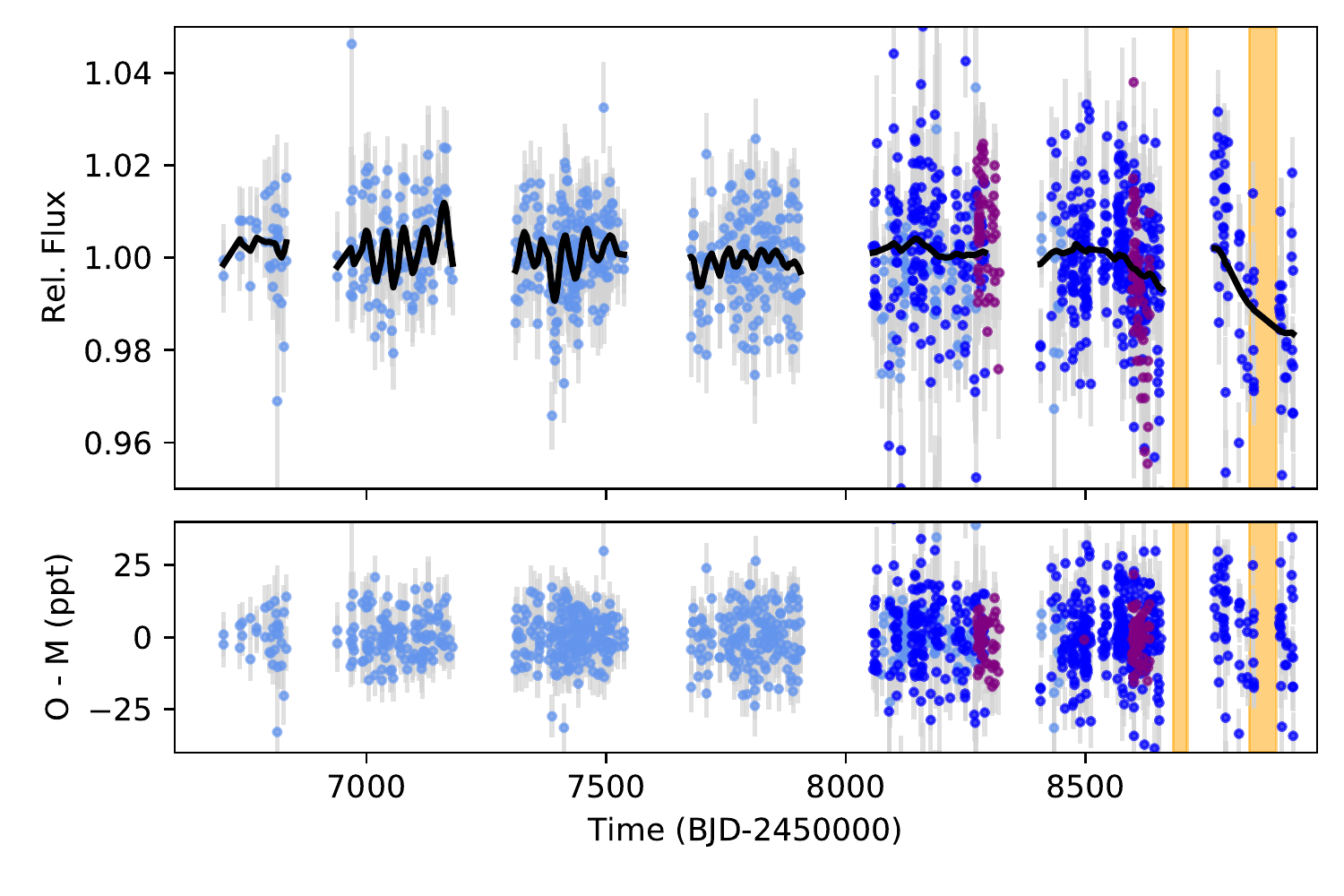}
    \caption{ASAS-SN ($V$ passband in light blue, $g'$ passband in dark blue) and NSVS (purple) long-term photometric monitoring modeled with a quasi-periodic GP kernel defined as in \citet{celerite}. 
    The time span of the {\em TESS} observations is shown in gold.
    ASAS-SN and NSVS fluxes were computed from the original magnitudes and normalized with their respective median.}
    \label{fig:ASAS_Prot}
\end{figure}

The low activity levels of TOI-1235 probably imply a slow rotation and that there may not be enough spot coverage to measure a rotation period.
Empirically, the measured limit on rotational velocity ($v \sin{i} <$ 2\,km\,s$^{-1}$) places a lower limit on $P_{\rm rot} / \sin{i} >$ 15.7\,d.
Given the estimate of the stellar radius in Table~\ref{tab:star}, the short-period transiting planet around such a low-mass stars indicates a low obliquity \citep{2017AJ....154..270W}, so that most probably, $\sin{i} \sim 1$ and therefore $P_{\rm rot} \gtrsim$ 16\,d.
On the other hand, from the $\log{R'_{\rm HK}}$-$P_{\rm rot}$ relation of \citet{2017A&A...600A..13A}, TOI-1235 has a most likely $P_{\rm rot}$ between 22\,d and 38\,d.
However, from \citet{Jeffers18}, the range of rotation periods that M dwarfs with $v \sin{i} <$ 2\,km\,s$^{-1}$ can have is between 10\,d and 150\,d.

To determine the actual rotational period of the star, we carried out different analyses of the available photometric data for TOI-1235. 
First, we employed the traditional periodogram analysis to search for significant peaks from the ASAS-SN $g'$- and $V$-band light curves. 
With the generalized Lomb-Scargle periodogram (GLS) of \citet{2009A&A...496..577Z}, we obtained a peak at 48.63 $\pm$ 0.08\,d above the 10\,\%~false-alarm probability (FAP) threshold for the combined light curve after subtracting an independent zero-point from each band. 
We explored the time parameter space between 10\,d and 1000\,d. 
Because the ASAS-SN light curves contained a significant number (15\,\%) of outlying data points because of flares and low S/Ns that might bias the previous GLS analysis, we repeated the GLS analysis after removing these deviant data from the two light curves in two steps: we first applied a $2\sigma$ and then a $1\sigma$ clipping algorithm. 
The new GLS periodogram of the resulting combined $g'$ and $V$ data looked different to the one of the original ASAS-SN data, as there were no significant peaks in the studied parameter space.
The highest peak near the 10\,\%~FAP level was located, but at a longer period of 136.9 $\pm$ 1.4\,d. 
The marginal amplitude of the cleaned ASAS-SN $g'$- and $V$-band light curve folded in phase with this long period was only 1.4\,mmag.

Next, we used a more sophisticated model and fit the ASAS-SN and NSVS photometry with a quasi-periodic Gaussian process (GP). 
In particular, we used the GP kernel introduced by \cite{celerite} of the form 
\begin{equation*}
k_{i,j}(\tau) = \frac{B}{2+C}e^{-\tau/L}\left[\cos \left(\frac{2\pi \tau}{P_\textnormal{rot}}\right) + (1+C)\right] \quad,
\end{equation*}
where $\tau = |t_{i} - t_{j}|$ is the time lag, $B$ and $C$ define the amplitude of the GP, $L$ is a timescale for the amplitude-modulation of the GP, and $P_{\rm rot}$ is the period of the quasi-periodic modulations. 
For the fit, we considered that each instrument and passband could have different values of $B$ and $C$, while $L$ and $P_{\rm rot}$ were left as common parameters. 
We considered wide uninformative priors for $B$, $C$ (log-uniform between $10^{-3}$\,ppm and $10^8$\,ppm), $L$ (log-uniform between $10^{2}$\,d and $10^8$\,d), $P_{\rm rot}$ (uniform between 10\,d and 300\,d), and instrumental jitter (log-uniform between 10\,ppm and $10^6$\,ppm). 
The fit was performed using \texttt{juliet} \citep[][see the next section for a full description of the algorithm]{juliet}, and the resulting fit is presented in Fig.~\ref{fig:ASAS_Prot}. 
The rotational period from the quasi-periodic GP analysis was found to be $P_\textnormal{rot} = 41.2^{+1.1}_{-1.2}$\,d, with an amplitude of about 10\,mmag during the time of highest stellar variability.

Finally, we took advantage of the {\em TESS} observations of TOI-1235 in three sectors spanning almost 210\,d.
We analyzed the light curve described in Sect~\ref{subsec:transit} and two light curves obtained from an optimized aperture (Gonz\'alez-Cuesta et~al.\ in prep.), in which we selected pixels with integrated flux above thresholds of 10\,e$^{-}$\,s$^{-1}$ and 20\,e$^{-}$\,s$^{-1}$, respectively. 
We then corrected the light curves for outliers and jumps, filled the gaps, concatenated the three sectors following \citet{2011MNRAS.414L...6G,2014A&A...568A..10G}, and removed the transits to make sure that they did not bias the results. 
Last, we applied our rotation pipeline \citep{2010A&A...511A..46M,2014A&A...572A..34G,2019ApJS..244...21S} with three different methods to search for a periodicity in the data: time-frequency analysis with wavelets \citep{1998BAMS...79...61T}, autocorrelation function \citep{2014ApJS..211...24M}, a and composite spectrum that is a combination of the the two previous methods \citep{2017A&A...605A.111C}. 
While different combinations of methods and light curves generally yielded somewhat different periodicities, signals in the range 32--42\,d were present in the time-frequency and composite spectrum analysis of the last two sectors.
The significance of the peaks in the autocorrelation function and composite spectrum was slightly below the criteria of in Section~3.3 of \citet{2017A&A...605A.111C} for establishing a reliable period (i.e., height of the peaks in the autocorrelation functions and composite function greater than or equal than 0.30 and 0.15, respectively).
To summarize, the GLS periodogram of the raw ASAS-SN data (although with a low significance), the quasi-periodic GP modeling of the combined ASAS-SN and NSVS data, and the s-BGLS analysis of the spectroscopic data (see Sect~\ref{subsec:GLS_rv}) all indicate a stellar origin of the $\sim$41.2\,d photometric signal, which suggests that this value might be the true rotation period of TOI-1235.

\subsection{Signals in spectroscopic data}\label{subsec:GLS_rv}

\begin{figure}[!ht]
    \centering
    \includegraphics[width=\hsize]{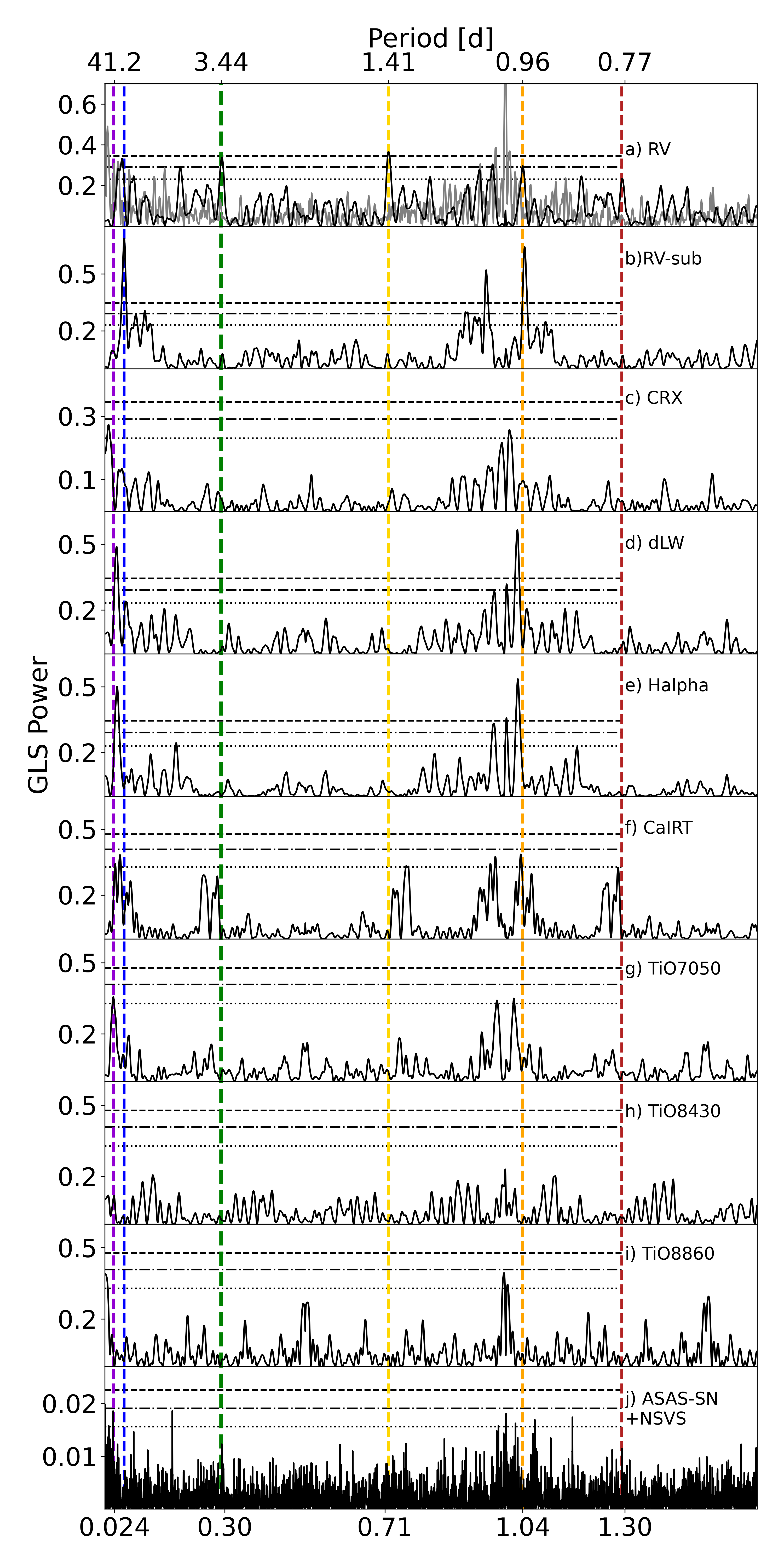}
    \caption{GLS periodograms of:
    ($a$) combined RVs from CARMENES VIS and HARPS-N (black) and the respective spectral window function (gray),
    ($b$) RV residuals after subtracting the planet signal,
    ($c$-$e$) combined CRX, dLW, and H$\alpha$ index from CARMENES VIS and HARPS-N, and
    ($f$-$i$) Ca~IRT, 
    TiO7050, TiO8430, and TiO8860 indices from CARMENES VIS alone, ($j$) combined ASAS-SN ($V$ and $g'$ bands), and NSVS data photometry.
    In all panels
    the vertical dashed lines indicate the periods of
    3.44\,d (thick green, planet) and the aliases of the orbital period (yellow, orange, and red),
    41.2\,d (violet, $P_{\rm rot}$ from the quasi-periodic GP analysis of the combined photometric data),
    20.6\,d  (blue, $\sim$\,$P_{\rm rot}$\,/2).
    The horizontal lines mark the theoretical FAP levels of 0.1\,\% (dashed), 1\,\% (dash-dotted), and 10\,\% (dotted).
    }    
    \label{fig:GLS_HARPS_CARMENES}
\end{figure}

We searched for periodic signals in the combined CARMENES VIS and HARPS-N RV data, which had the lowest median RV uncertainties, by computing GLS periodograms, as illustrated by Fig.~\ref{fig:GLS_HARPS_CARMENES}. 
A signal corresponding to the transits in the \textit{TESS} light curve was significantly detected in the RVs at $P_{\rm b}$~=~3.44\,d (FAP $\sim$1\,\%; panel $a$) and its aliases at 1.41\,d, 0.96\,d, and 0.77\,d.
However, we also found an additional signal at $P \approx$~20.6\,d, at about half the most likely stellar rotation period.
After removing the planetary signal, the 20.6\,d signal and its aliases still remained with an FAP $\gtrsim$ 0.1\,\% (panel $b$).

To understand the origin of the 20.6\,d signal, we searched for additional peaks in the periodograms of the activity indicators CRX, dLW, and H$\alpha$ derived from the individual CARMENES and HARPS-N spectra (panels $c$-$e$), and Ca~IRTa (panel $f$), and the titanium oxide indices that quantify the strengths of the TiO $\gamma$, $\epsilon$, and $\delta$ absorption band heads at 7050\,{\AA}, 8430\,{\AA}, and 8860\,{\AA} (panels $g$-$i$), respectively, from the CARMENES spectra alone \citep{2018A&A...609A..12Z, Schf19}. 
The activity indices and their uncertainties are listed in Table~\ref{tab:RV_Activity_all}.
Except for daily aliases, the highest peaks in the dLW, H$\alpha$, Ca~IRTa, and TiO7050 periodograms are at about $P~\approx$ 32--47\,d, which adds further credence to $P_{\rm rot} \approx$ 41.2\,d as inferred in Section~\ref{subsubsec:Prot}.  
All these indicators track different features in the stellar atmosphere, and our spectra cover only slightly more than two periods, therefore it is plausible that they do not yield exactly the same periods. 
We also detected the 20.6\,d signal in the dLW series, which supports the notion that this signal is also related to stellar activity. 
As expected for an early-type M dwarf, the TiO8430 and TiO8860 indices showed no significant signals.

\begin{figure}
\centering
\includegraphics[width=0.24\textwidth]{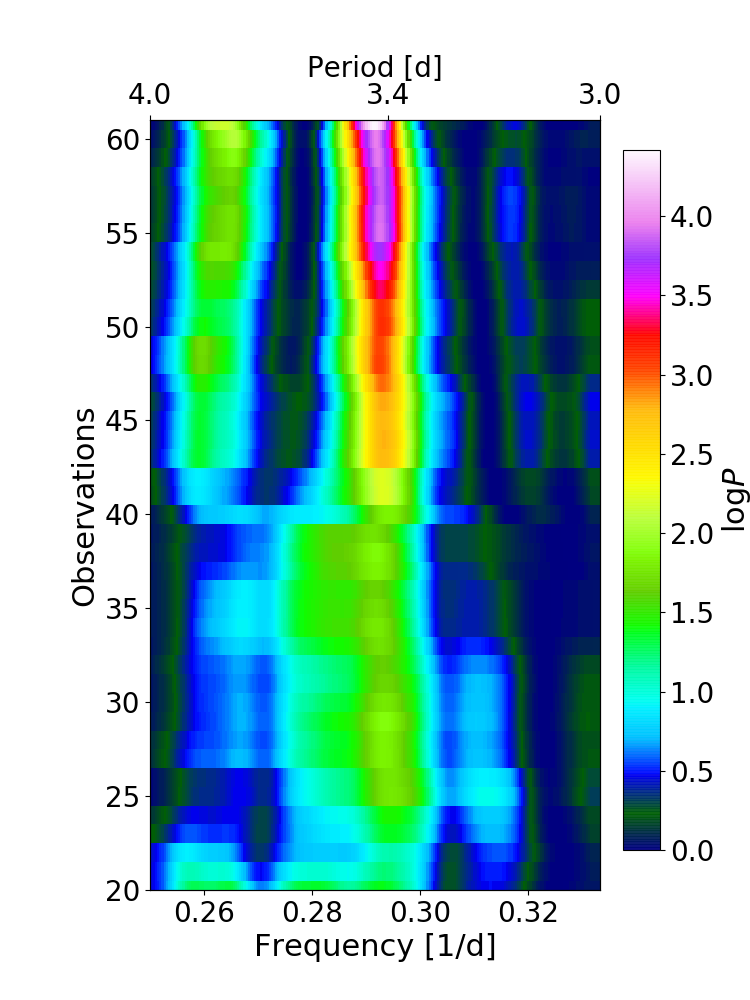}
\includegraphics[width=0.24\textwidth]{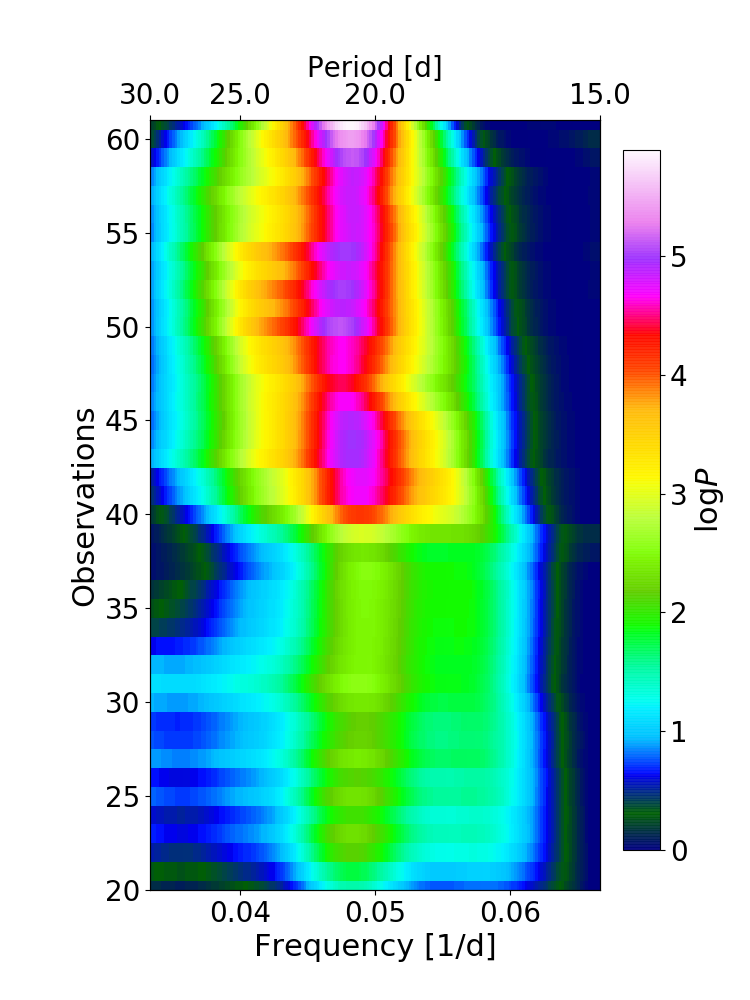}
\caption{Evolution of the s-BGLS periodogram of the CARMENES and HARPS-N RV data of TOI-1235 around the 3.44\,d signal of the transiting planet ({\em left}) and around the 20.6\,d activity signal after subtracting the planet signal ({\em right}). The number of data points included in the computation of the periodogram increases from bottom to top.}
\label{fig: sBGLS_TOI1235}
\end{figure}

We used the stacked Bayesian generalized Lomb-Scargle periodogram \citep[s-BGLS;][]{BGLS2015} with the normalization of \cite{sBGLS2017} to verify whether the 20.6\,d signal was coherent over the whole observational time baseline of CARMENES VIS and HARPS-N. 
In Fig.~\ref{fig: sBGLS_TOI1235} we display s-BGLS periodograms of the raw RV data around 3.44\,d, and of the RV data, after subtracting a sinusoid at the transiting planet period, around the 20.6\,d signal. 

This signal showed a first probability maximum after around 44 observations (BJD $\sim$ 2458663) and thereafter decreased for some time. 
This incoherence is characteristic of a nonplanetary origin of the signal \citep{sBGLS2017}. 
The s-BGLS of the 3.44\,d signal, on the other hand, showed a monotonically increasing probability, as expected for a Keplerian signal. 

Last, we measured the Pearson $r$, Student $t$, and Fisher $z$ correlation coefficients between the temporal series of RV and the activity indicators CRX, dLW, H$\alpha$, Ca~IRTa, TIO7050, and $S_{\rm MWO}$, and we did not see any intrinsic correlation between RV and activity as in \citet{2020AJ....159..160G}.
In particular, we determined absolute values of $r$ and $z$ below 0.006 and of $t$ above 0.7, respectively, for all relations except for RV versus dLW, which was in any case weakly anticorrelated.

\subsection{Joint fit} \label{subsec:joint}

\begin{figure*}
    \centering
    \includegraphics[width=\textwidth]{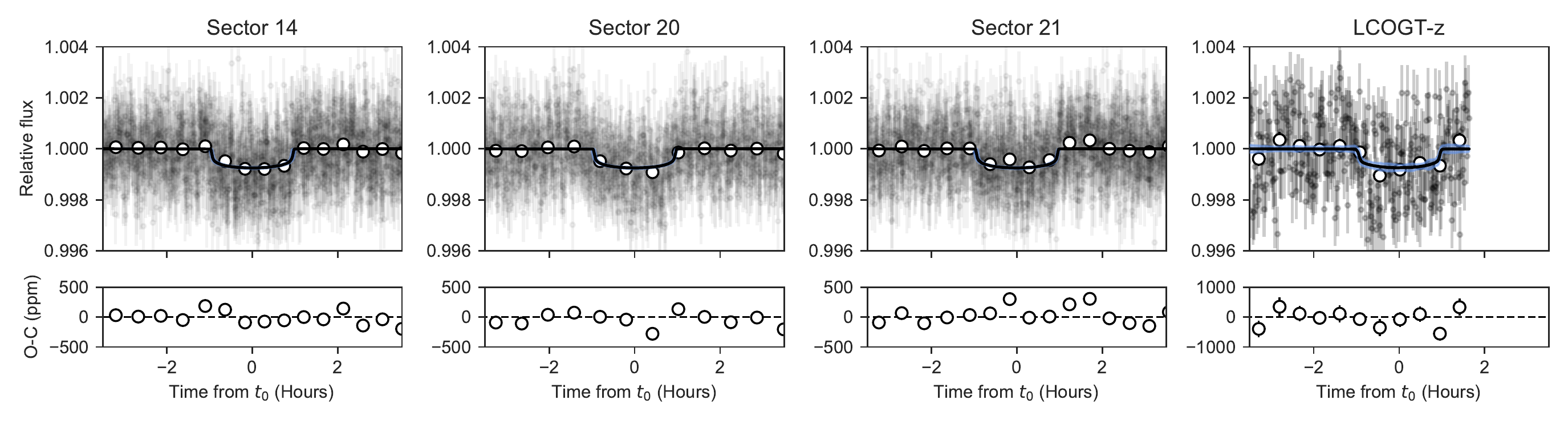}
    \includegraphics[width=\textwidth]{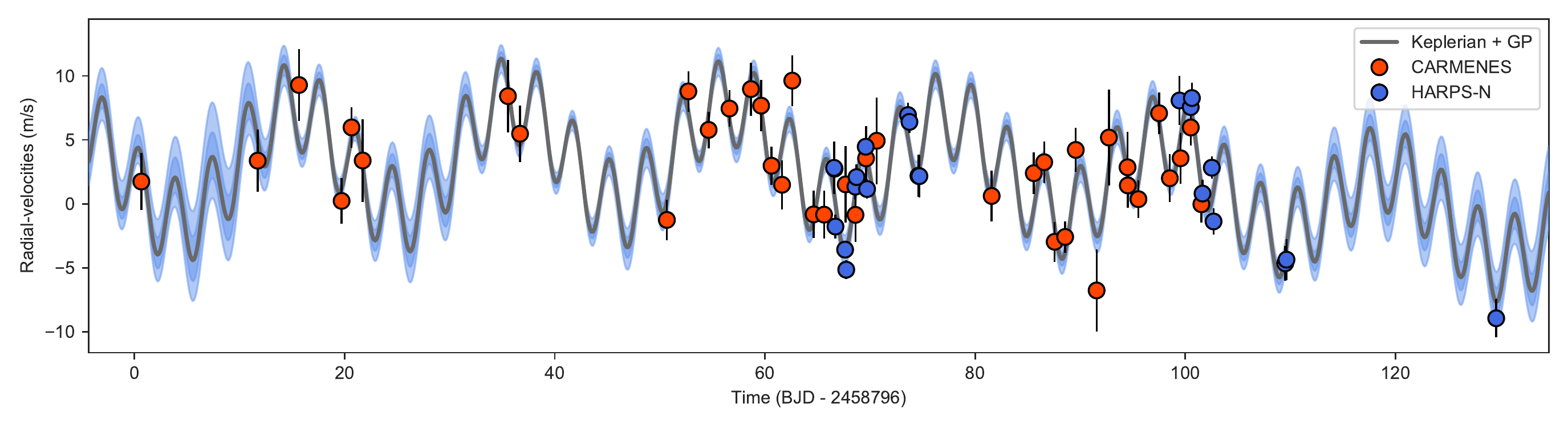}
    \caption{Joint fit results.
    {\em Top panels:}
    Phase-folded light curves of {\em TESS}, sectors 14, 20, and 21, and LCOGT, from left to right, and their residuals.
    White circles are binned data (shown only for reference; data used to fit the model were the unbinned points), black curves are the best-fit models, and blue areas are the 68\,\% credibility bands.
    {\em Bottom panel:}
    CARMENES VIS (orange) and HARPS-N (blue) RVs.
    The gray curve is the median best-fit \texttt{juliet} model, and the light and dark blue areas are its 68\,\% and 95\,\% credibility bands.
    }
    \label{fig:joint-fit-photo}
\end{figure*}

To obtain precise parameters of the TOI-1235 system, we performed a joint analysis of the {\it TESS} and LCOGT photometry and CARMENES VIS and HARPS-N RV data using \texttt{juliet}.
The model that we selected for our RV joint fit analysis was one composed of a circular Keplerian orbit for the transiting planet plus a quasi-periodic GP that we used to model the 20.6\,d signal observed in the RVs; we have  discussed this in previous sections.
However, we also computed models of a circular orbit, an eccentric orbit, a circular orbit plus a sinusoid, an eccentric orbit plus a sinusoid, an eccentric orbit plus a GP, and two circular orbits.
The two best-fit models, judged by their log-evidences, were a two-planet model and a one-circular-planet model combined with a GP to fit the 20.6\,d signal.
The star-planet tidal locking and consequent circularization of the planet orbit following the theoretical predictions of \citet{2017CeMDA.129..509B}, for instance, support both models with eccentricity fixed to zero (see also Fig.~2 in \citealt{2019ApJ...887..261M}). 
However, the difference between their log-evidence was $\Delta \ln{\mathcal{Z}} < 2$, which made the two models indistinguishable if they were equally likely a priori. 
The two models gave almost identical constraints on the properties of the transiting exoplanet.
The analyses of the activity indices and photometric data, however, gave a higher prior weight to the stellar activity model, and we therefore decided to use a GP, which is typically better at modeling stellar activity than a simple sinusoid, as our final model to account for the 20.6\,d signal. 
In our analysis we used the exp-sine-squared kernel for the GP, which is a very common kernel to model stellar activity signatures in the literature \citep[see, e.g.,][and references therein]{nava2020}, and which is of the form  
\begin{eqnarray*}
k_{i,j}(\tau) = \sigma^2_{\rm GP} \exp\left(-\alpha \tau^2 - \Gamma \sin^2\left[\frac{\pi \tau}{P_{\rm rot}}\right]\right).
\end{eqnarray*}
For the transit modeling, \texttt{juliet} uses the \texttt{batman} package \citep{batman}. 
To parameterize the limb-darkening effect in the 
\textit{TESS} photometry, we employed the efficient, uninformative sampling scheme of \cite{kipping2013} and a quadratic law.
We used a common set of limb-darkening coefficients across the three \textit{TESS} sectors.
In the LCOGT light-curve analysis, we instead used a linear law to parameterize the limb-darkening effect, as a more complex law was not warranted given the precision of the data, as explained by \citet{EJ16}.
We used the \cite{Espinoza18} parameterization to explore the full physically plausible parameter space for the planet-to-star radius ratio, $R_{\rm p}/R_\star$, and impact parameter, $b$. 
Finally, we used a white-noise-only fit for the \textit{TESS} photometry, as an analysis using a GP on the photometry returned a log-evidence that was indistinguishable from the one of a white-noise model. 
For the LCOGT photometry, on the other hand, we used a linear model to detrend the data, with airmass and pixel position of the target as regressors.
The selected priors for our joint fit are presented in Table \ref{tab:priors}. 

\begin{figure}
    \centering
    \includegraphics[width=\hsize]{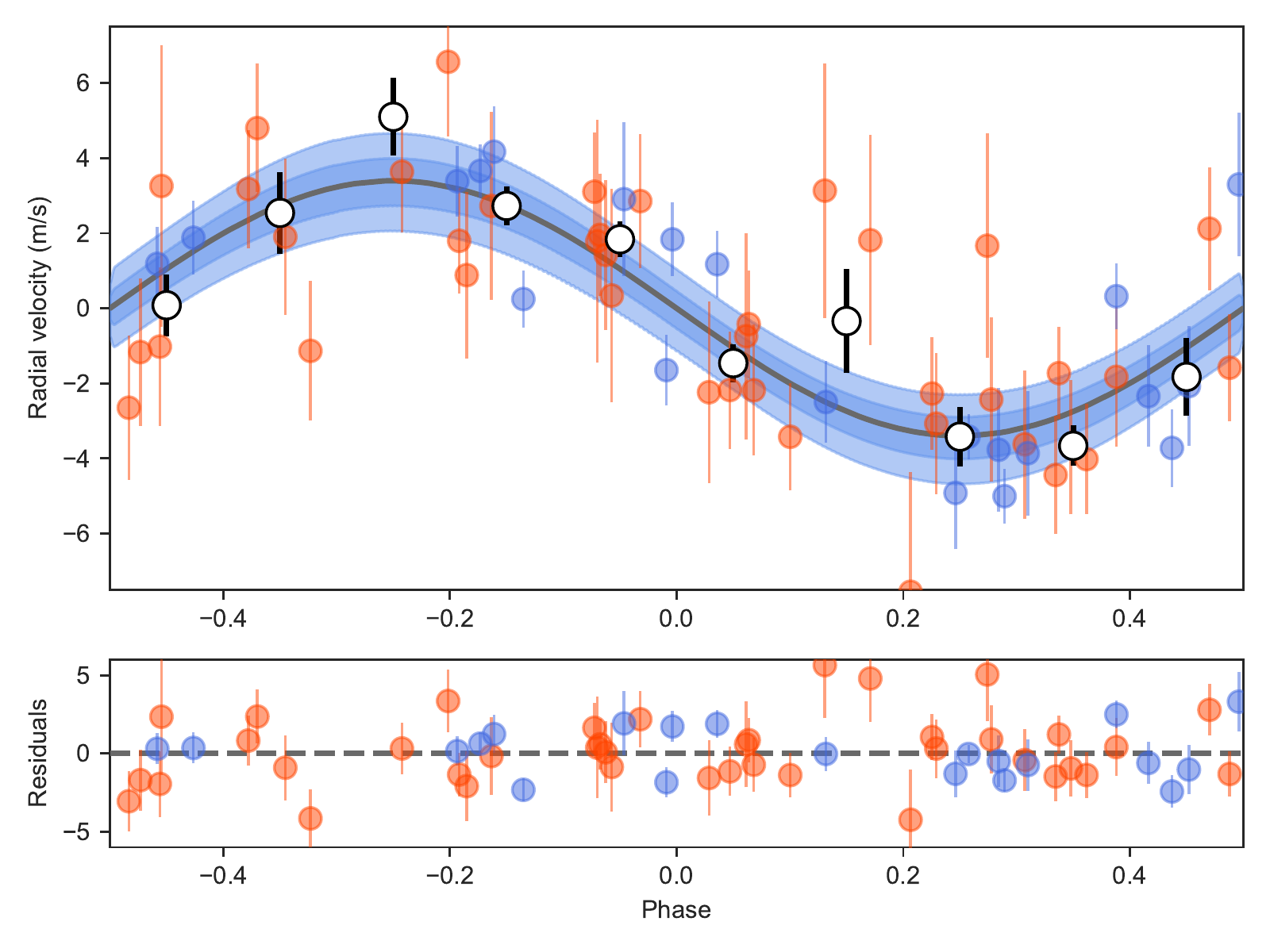}
    \caption{Phase-folded RVs for TOI-1235 without the GP component.
    Orange circles are CARMENES VIS data, blue circles are HARPS-N data, white points are binned data for reference.
    The gray curve is the median best-fit \texttt{juliet} model, and the light and dark blue areas are its 68\,\% and 95\,\% credibility bands.} 
    \label{fig:phased-rvs}
\end{figure}

    \begin{table}
    \centering
    \caption{Posterior parameters of the \texttt{juliet} joint fit for TOI-1235~b.}
    \label{tab:posteriors}
    \begin{tabular}{lc} 
        \hline
        \hline
        \noalign{\smallskip}
        Parameter$^{a}$ & TOI-1235~b \\
        \noalign{\smallskip}
        \hline
        \noalign{\smallskip}
        \multicolumn{2}{c}{\it Stellar parameters} \\[0.1cm]
        \noalign{\smallskip}
        $\rho_\star$ ($\mathrm{g\,cm\,^{-3}}$)& $3.74^{+0.30}_{-0.31}$\\[0.1 cm]
        \noalign{\smallskip}
        \multicolumn{2}{c}{\it Planet parameters} \\[0.1cm]
        \noalign{\smallskip}
        $P$ (d)                              & $ 3.444717^{+ 0.000040}_{- 0.000042}$\\[0.1 cm]
        $t_0$ (BJD)                & $2458683.6155^{+ 0.0017}_{- 0.0015}$\\[0.1 cm]
        $a/R_\star$                                  & $13.29^{+ 0.34}_{- 0.38}$ \\[0.1 cm]
        $p = R_{\rm p}/R_\star$                                  & $0.02508^{+ 0.00084}_{- 0.00085}$ \\[0.1 cm]
        $b = (a/R_\star)\cos i_{\rm p}$                                  & $0.25^{+ 0.12}_{- 0.14}$ \\[0.1 cm]
        $i_{\rm p}$ (deg)   & $88.90^{+ 0.62}_{- 0.57}$ \\[0.1 cm]
        $r_1$                                & $0.500^{+ 0.081}_{- 0.097}$ \\[0.1 cm]
        $r_2$                                & $0.02506^{+ 0.00083}_{- 0.00085}$\\[0.1 cm]
        $K$ ($\mathrm{m\,s^{-1}}$)           & $3.40^{+ 0.35}_{- 0.34}$\\[0.1 cm]
        \noalign{\smallskip}
        \multicolumn{2}{c}{\it Photometry parameters} \\[0.1cm]
        \noalign{\smallskip}
        $M_{\mathrm{TESS,S14}}$ ($10^{-6}$)   & $-31.0^{+ 8.5}_{- 8.3}$\\[0.1 cm]
        $M_{\mathrm{TESS,S20}}$ ($10^{-6}$)   & $-17.0^{+ 8.3}_{- 8.2}$\\[0.1 cm]
        $M_{\mathrm{TESS,S21}}$ ($10^{-6}$)   & $-24.0^{+ 8.0}_{- 8.0}$\\[0.1 cm]
        $\sigma_{\mathrm{TESS,S14}}$ (ppm)     & $1.9^{+ 10.5}_{- 1.6}$\\[0.1 cm]
        $\sigma_{\mathrm{TESS,S20}}$ (ppm)     & $1.9^{+ 8.2}_{- 1.6}$\\[0.1 cm]
        $\sigma_{\mathrm{TESS,S21}}$ (ppm)     & $1.5^{+ 7.8}_{- 1.3}$\\[0.1 cm]   
        $q_{1,\mathrm{TESS}}$               & $0.42^{+ 0.32}_{- 0.25}$\\[0.1 cm]
        $q_{2,\mathrm{TESS}}$               & $0.31^{+ 0.30}_{- 0.20}$\\[0.1 cm]
        $M_{\mathrm{LCO}}$ ($10^{-6}$)      & $-257^{+ 84}_{- 86}$\\[0.1 cm]
        $\sigma_{\mathrm{LCO}}$ (ppm)        & $970^{+ 82}_{- 83}$\\[0.1 cm] 
        $q_{1,\mathrm{LCO}}$                 & $0.49^{+ 0.30}_{- 0.30}$\\[0.1 cm]
        $\theta_{0,\mathrm{LCO}}$ ($10^{-6}$)      & $-10^{+ 11}_{- 11}$\\[0.1 cm]   
        $\theta_{1,\mathrm{LCO}}$ ($10^{-6}$)      & $-49^{+ 11}_{- 11}$\\[0.1 cm]          
        \noalign{\smallskip}
        \multicolumn{2}{c}{\it RV parameters}\\[0.1cm]
        \noalign{\smallskip}
        $\gamma_{\mathrm{CARMENES}}$ ($\mathrm{m\,s^{-1}}$)       & $-3.0^{+ 4.6}_{- 4.3}$\\[0.1 cm]
        $\sigma_{\mathrm{CARMENES}}$ ($\mathrm{m\,s^{-1}}$)    & $0.17^{+ 0.61}_{- 0.14}$\\[0.1cm]
        $\gamma_{\mathrm{HARPS-N}}$ ($\mathrm{m\,s^{-1}}$)        & $3.8^{+ 4.6}_{- 4.2}$\\[0.1 cm]
        $\sigma_{\mathrm{HARPS-N}}$ ($\mathrm{m\,s^{-1}}$)     & $1.29^{+ 0.43}_{- 0.37}$\\[0.1 cm]
        \noalign{\smallskip}
        \multicolumn{2}{c}{\it GP hyperparameters} \\
        \noalign{\smallskip}
        $\sigma_\mathrm{GP,RV}$ ($\mathrm{m\,s^{-1}}$)              & $12.3^{+ 17.9}_{- 6.3}$\\[0.1 cm]
        $\alpha_\mathrm{GP,RV}$ ($10^{-6}\,\mathrm{d^{-2}}$)        & $74^{+ 127}_{- 50}$\\[0.1 cm]
        $\Gamma_\mathrm{GP,RV}$                           & $0.084^{+ 0.251}_{ -0.068}$\\[0.1 cm]
        $P_\mathrm{rot;GP,RV}$ (d)                              & $20.93^{+ 0.56}_{- 0.52}$\\[0.1 cm]
        \noalign{\smallskip}
        \hline
    \end{tabular}
    \tablefoot{
        \tablefoottext{a}{Priors and descriptions for each parameter are in Table~\ref{tab:priors}. Error bars denote the 68\,\% posterior credibility intervals.}}
\end{table}

As illustrated by the posterior parameters of our joint fit presented in Table~\ref{tab:posteriors} and the resulting RV model presented in Fig.~\ref{fig:joint-fit-photo}, the maximum a posteriori of the GP periodic component, $P_\mathrm{rot;GP,RV}$, is about 20.9\,d, in agreement with the signal observed in the GLS periodogram of the RVs (Fig.~\ref{fig:GLS_HARPS_CARMENES}).
This is almost exactly half the period derived from the long-term photometric monitoring discussed in previous sections, which means that a rotating spotted stellar surface is the most plausible cause of these variations. 
Consequently, we performed joint fits using the period observed in the photometry of $P_{\rm rot} = 41.2^{+1.2}_{-1.5}$\,d as a prior, and the results were almost identical regarding the properties of the transiting planet to the ones presented in Table~\ref{tab:posteriors}. 
Therefore our model is marginalized properly over the possible different scenarios on the stellar surface in terms of stellar activity. 
As shown in Fig.~\ref{fig:phased-rvs} and Table~\ref{tab:posteriors}, we attained a $10\sigma$ detection of the planetary RV semiamplitude.

We also performed two additional fits of the the lower-precision CARMENES NIR and iSHELL RVs (Sects.~\ref{subsec:carmenes} and~\ref{subsec:iSHELL}).
We set all the ephemeris priors to those found in the joint fit, including the planet $P$ and $t_0$, with and without stellar rotation period and timescale of the GP. 
All other parameters were free to vary around the entire parameter space.
The two models, with and without GP, were indistinguishable based on their log-evidences ($\Delta \ln{\mathcal{Z}} < 1$).
Interestingly, the GP amplitude in the first model was consistent with zero, which supports the nonplanetary origin of the $\sim$41.2\,d period, as argued above.
The new NIR RV analysis yielded a lower statistical precision in model parameter recovery than VIS RVs, but the new recovered planet-semiamplitude $K_{\rm NIR}~=~2.8\pm1.4$\,m\,s$^{-1}$ was consistent within $1\sigma$ with that listed in Table~\ref{tab:posteriors}.   
In addition, the NIR RVs, taken at independent wavelengths and in the case of iSHELL with a different facility, help validate the system and instrument performance.

To summarize, the TOI-1235 system consists of a relatively weakly active M dwarf with at least one super-Earth-like planet, namely TOI-1235\,b (see~Table~\ref{tab:derivedparams}), with a mass of $M_{\rm p}$~=~$5.9 ^{+ 0.6}_{- 0.6}$\,$M_\oplus$ and radius of $R_{\rm p}$~=~$1.69^{+0.08}_{-0.08}$\,$R_\oplus$ on a circular orbit with a period of 3.44\,d.
We also derived a bulk density of $\rho_{\rm p}$~=~$6.7^{+ 1.3}_{-1.1}$\,g\,cm$^{-3}$ and an equilibrium temperature, assuming a zero albedo, of $T_\textnormal{eq}$~=~$776\pm13$\,K, which is slightly hotter than the mean surface temperature of Venus.

\begin{table}
    \centering
    \caption{Derived planetary parameters for TOI-1235~b.}
    \label{tab:derivedparams}
    \begin{tabular}{lc} 
        \hline
        \hline
        \noalign{\smallskip}
        Parameter\tablefootmark{a} & TOI-1235~b  \\
        \noalign{\smallskip}
        \hline
        \noalign{\smallskip}
        \multicolumn{2}{c}{\it Derived transit parameters} \\[0.1cm]
        \noalign{\smallskip}
        $u_1$\tablefootmark{b}             & $0.38 ^{+ 0.30}_{- 0.24}$    \\[0.1 cm]
        $u_2$\tablefootmark{b}             & $0.22 ^{+ 0.35}_{- 0.32}$     \\[0.1 cm]
        $t_T$ (h)                & $2.094 ^{+ 0.126}_{- 0.086}$      \\[0.1 cm]
        \noalign{\smallskip}
        \multicolumn{2}{c}{\it Derived physical parameters} \\[0.1cm]
        \noalign{\smallskip}
        $M_{\rm p}$ ($M_\oplus$)    & $5.90 ^{+ 0.62}_{- 0.61}$    \\[0.1 cm]
        $R_{\rm p}$ ($R_\oplus$)        & $1.694 ^{+ 0.080}_{- 0.077}$   \\[0.1 cm]
        $\rho_{\rm p}$ (g cm$^{-3}$)             & $6.7 ^{+ 1.3}_{- 1.1}$      \\[0.1 cm]
        $g_{\rm p}$ (m s$^{-2}$)                 & $20.1 ^{+ 3.0}_{- 2.7}$     \\[0.1 cm]
        $a_{\rm p}$ (au)                         & $0.03826 ^{+ 0.00048}_{- 0.00049}$  \\[0.1 cm]
        $T_\textnormal{eq}$ (K)\tablefootmark{c}          & $775 ^{+ 13}_{- 13}$    \\[0.1 cm]
        $S$ ($S_\oplus$)            & $60.3 ^{+ 1.6}_{- 1.5}$  \\[0.1 cm]
        \noalign{\smallskip}
        \hline
    \end{tabular}
    \tablefoot{
      \tablefoottext{a}{Parameters obtained with the posterior values from Table~\ref{tab:posteriors}, $t_T$=Transit duration, from first contact to fourth contact.
      Error bars denote the 68\,\% posterior credibility intervals.}
      \tablefoottext{b}{Derived from the {\it TESS} light curve.}
      \tablefoottext{c}{The equilibrium temperature was calculated assuming zero Bond albedo.}
      }
\end{table}

\subsection{Search for transit depth and time variations}\label{subsec:TTV}

{\em TESS} observed TOI-1235 in three sectors and covered 22 transits of TOI-1235\,b. This allowed us to assess the presence of transit-timing variations (TTVs) and transit depth variations. 
We carried out a search for TTVs using the \texttt{batman} package and fit each transit individually. We only left transit times and transit depth as free parameters, and fixed the remaining parameters to the values obtained in the joint analysis in Sect.~\ref{subsec:joint}.
The best-fit parameters and associated uncertainties in our fitting procedure were derived using a Markov chain Monte Carlo analysis implemented in the \texttt{emcee} python package \citep{Foreman-Mackey-13}.
We found an indication of a periodic TTV signal with a semiamplitude of about 4\,min. 
Using the GLS of the observed TTV signal, we found that the observed TTVs had a periodicity of 25.3$\pm$0.2\,d, which might indicate the presence of a second nontransiting planet in the system \citep{2005Sci...307.1288H}. 
However, a TTV signal with this amplitude might also easily be generated by the stellar activity \citep[e.g.,][]{Oshagh-13}, and the period was consistent with our previous analyses of the stellar rotation. 
We also searched for trends in the derived transit depths, and found that individual depths agreed within $1\sigma$ with the depth derived from the combined analysis. 

\section{Discussion} \label{sec:discussion}

\begin{figure*}
    \centering
    \includegraphics[width=\hsize]{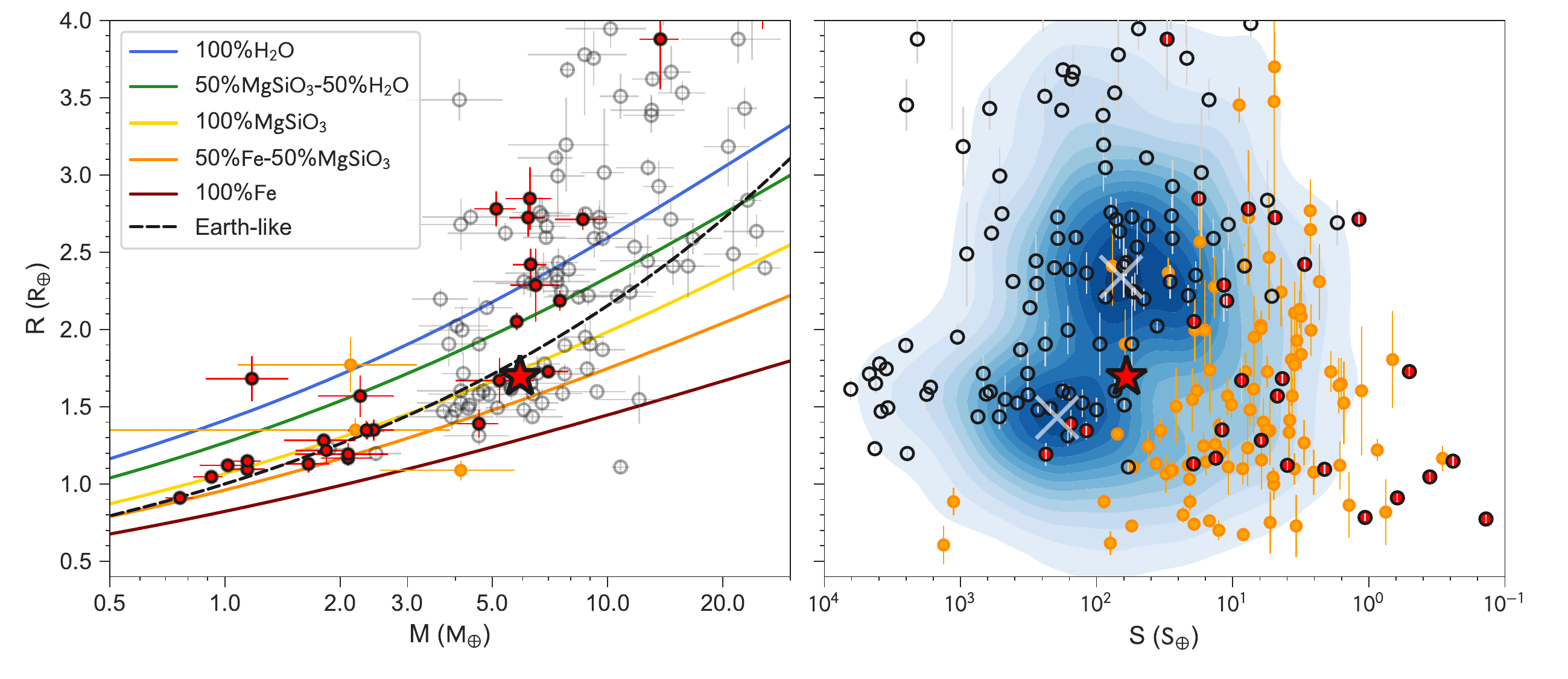}
    \caption{Mass-radius ({\em left}) and insolation-radius ({\em right}) diagrams in Earth units.
    In the two panels, 
    open circles are transiting planets around F-, G-, and K-type stars with mass and radius measurement better than 30\,\% from the TEPCat database of well-characterized planets \citep{2011MNRAS.417.2166S},
    filled red circles are planets around M dwarfs with mass and radius measurement, filled yellow circles are
    planets around M dwarfs with mass determinations lower than 30\,\% or without mass constraints at all ({\em right} panel only),
    and the red star is TOI-1235\,b, whose radius and mass are determined with accuracies of 5\,\% and 10\,\%, respectively.
    In the {\em left} panel, the color lines are the theoretical $R$-$M$ models of \citet{2016ApJ...819..127Z}, 
    and the three planets with mass determinations lower than 30\,\% are \object{K2--3}\,b, \object{BD--17~588}A\,b, and \object{LHS~1815}\,b.
    In the {\em right} panel, we plot
    the $R$-$S$ point density of all the known confirmed transiting planets with contours, and mini-Neptunes and super-Earths density maxima with white crosses.
    The M dwarf without mass determination in the radius gap is \object{K2--104}\,b \citep{2017AJ....153...64M}, a planet around an active star in the Praesepe cluster that is fainter by 5\,mag in $V$ than TOI-1235.} 
    \label{fig:mass_insolation_radius}
\end{figure*}

Our 61 RV measurements yield a planetary mass for TOI-1235\,b with an uncertainty of about 10\,\%, and the \textit{TESS} and LCOGT light curves constrain the planetary radius at a level of about 5\,\% uncertainty. This means that TOI-1235\,b belongs to the select group of terrestrial planets with a well-determined bulk density. The population with measurements better than 30\,\% is shown in the mass-radius diagram of Fig.~\ref{fig:mass_insolation_radius}. 
The comparison of TOI-1235\,b with theoretical models of \citet{2016ApJ...819..127Z,Zeng2019} is consistent with a rocky MgSiO$_3$-dominated composition with a bulk density slightly higher than that of Earth. This classifies it as a super-Earth planet. 

Based again on the mass and radius relationships from \citet{2016ApJ...819..127Z}, the best fit results in an iron core mass fraction of CRF~=~0.10$^{+0.38}_{-0.10}$, but the planet is also consistent with an Earth-like bulk composition (CRF~$\approx$~0.4--0.6).
Furthermore, using {\tt Hardcore} \citep{2018MNRAS.476.2613S} and our $R$ and $M$, the marginal core ratio fraction, CRF$_{\rm marg}$, is 0.53$\pm$0.20, similar to the Earth's true CRF value of 0.55.

Like many other transiting terrestrial and sub-Neptune planets, TOI-1235\,b is on a fairly irradiated orbit and therefore may have been strongly sculpted by extreme atmospheric escape due to XUV-driven photoevaporation \citep[e.g.,][]{Lopez2013,owen2013} or core-powered mass loss \citep[e.g.,][]{wu2019,gupta2020}. 
Because of its expected low envelope mass fraction, the
required binding energy makes this explanation difficult for TOI-1235\,b, but using the escape scaling relations from \citet{Lopez2013}, we found that this planet lies right at the boundary of where escape evolution is likely to play a significant role in removing primordial H/He gaseous envelopes.

As described in Sect.~\ref{sec:intro} and illustrated by the insolation-radius diagram in Fig.~\ref{fig:mass_insolation_radius},
the growing exoplanet statistics has revealed a gap in the radius distribution of planets slightly larger than Earth \citep{Fulton17}. 
Rocky super-Earth planets of up to $\sim$1.5\,$R_\oplus$ are relatively common, as are gaseous mini-Neptunes in the range of 2--4\,$R_\oplus$, but only a few planets have been detected with a radius inside this gap \citep{2019ApJ...876L..24G}. 
Using the location of the radius valley as determined by \cite{vaneylen2018}, that is, $\log{R} = m \log{P} + a$ with $m = -0.09^{+0.02}_{-0.04}$ and $a = 0.37^{+0.04}_{-0.02}$, we determine the predicted location of the radius valley at the orbital period of TOI-1235\,b. 
We find that for $P = 3.44$\,d, the radius valley is located at $R = 2.1 \pm 0.2\,R_\oplus$. Therefore and according to this definition, TOI-1235\,b, which has a radius $ R = 1.69^{+0.08}_{-0.07}\,R_\oplus$, would be located near the lower edge of the radius valley. 
Its rocky composition is indeed consistent with the planet having lost its atmosphere, as expected for planets below the radius valley \citep[e.g.,][]{owen2013}. 

However, the location of the radius gap as determined by \cite{vaneylen2018} was based on F-, G-, and K-type stars, whereas TOI-1235\,b orbits an M dwarf star. 
Whether these same boundaries apply to M dwarfs (and whether the gap actually exists for planets around M dwarfs) has been the subject of several recent studies \citep{Zeng2017, Fulton18, Hirano2018}. Following \cite{Zeng2017}, for example, who used all of the \textit{Kepler} planet candidates, the radius and stellar irradiation level of TOI-1235\,b place it exactly in the gap for early-M dwarfs (located at about 1.7\,$R_\oplus$ for an irradiation of 60\,$S_\oplus$ in that work). On the other hand, when we extrapolate from the sample of \cite{Fulton18}, who focused on F-, G-, and K-type stars with precise stellar parameters and on stars that host validated \textit{Kepler} exoplanets, we reach a similar conclusion. Finally, using the sample of \cite{Hirano2018}, who focused only on low-mass stars hosting validated small planets unveiled by K2 and {\em Kepler}, we would locate TOI-1235\,b in the gap, but the data in that sample (arguably more suitable for a proper comparison with the stellar properties of TOI-1235) were unable to track a proper stellar irradiation versus radius dependence of the gap. Therefore our measurements of the bulk composition of TOI-1235\,b, consistent with the planet having lost its atmosphere, place a strong constraint on any interpretation regarding the radius gap for M dwarfs at the irradiation levels received by TOI-1235\,b.
If atmospheric loss is indeed the correct physical interpretation for the radius gap, and if it applies to M dwarfs at the period or stellar irradiation level of TOI-1235\,b, the gap for early-type M dwarfs has to be either at or above 1.7\,$R_\oplus$.

\section{Conclusions} \label{sec:conclusions}

We confirmed that TOI-1235\,b is a transiting super-Earth planet around an M0.5\,V star, observed in sectors 14, 20, and 21 of the \textit{TESS} mission. 
We collected CARMENES and HARPS-N spectroscopic data, from which we confirmed the planetary nature of the transit signal detected by \textit{TESS}. 
Further support for the planetary interpretation came from our LCOGT photometric data during one transit, as well as from lucky and speckle imaging.
From the joint analysis of all the data, we derived the following parameters for TOI-1235\,b: mass of $M_{\rm p}$~=~5.9$\pm$0.6\,$M_{\oplus}$, radius of $R_{\rm p}$=1.69$\pm$0.08\,$R_{\oplus}$, and density of  $\rho_{\rm p}$~=~$6.7 ^{+ 1.7}_{- 1.1}$\,g\,cm$^{-3}$. 

A comparison of the physical properties of TOI-1235\,b with compositional models reveals the planet to be a rocky super-Earth, with a bulk density only slightly higher than that of Earth. Although the location (and existence) of a radius gap for exoplanets around M-dwarfs is still debated, the radius and irradiation level of TOI-1235\,b place it at the radius gap according to various suggestions of its location in the literature for these small low-mass stars. If the gap indeed exists for M-dwarfs, the bulk properties of TOI-1235\,b, which make it consistent with having lost its atmosphere, constrain the gap to be at or above the planetary radius of TOI-1235\,b, that is, $\sim 1.7\,R_\oplus$ at its irradiation level ($\sim 60\,S_\oplus$). These findings help to better constrain the dependence of the gap location on stellar type and irradiation, and thus to understand its origin. Finally, the brightness of TOI-1235 ($V \approx$ 11.5\,mag) makes this planet an accessible and very interesting object for further studies of planet formation and atmospheric evolution.

\begin{acknowledgements}
  CARMENES is an instrument for the Centro Astron\'omico Hispano-Alem\'an de
  Calar Alto (CAHA, Almer\'{\i}a, Spain). 
  CARMENES is funded by the German Max-Planck-Gesellschaft (MPG), 
  the Spanish Consejo Superior de Investigaciones Cient\'{\i}ficas (CSIC),
  the European Union through FEDER/ERF FICTS-2011-02 funds, 
  and the members of the CARMENES Consortium 
  (Max-Planck-Institut f\"ur Astronomie,
  Instituto de Astrof\'{\i}sica de Andaluc\'{\i}a,
  Landessternwarte K\"onigstuhl,
  Institut de Ci\`encies de l'Espai,
  Institut f\"ur Astrophysik G\"ottingen,
  Universidad Complutense de Madrid,
  Th\"uringer Landessternwarte Tautenburg,
  Instituto de Astrof\'{\i}sica de Canarias,
  Hamburger Sternwarte,
  Centro de Astrobiolog\'{\i}a and
  Centro Astron\'omico Hispano-Alem\'an), 
  with additional contributions by the Spanish Ministry of Economy, 
  the German Science Foundation through the Major Research Instrumentation 
    Programme and DFG Research Unit FOR2544 ``Blue Planets around Red Stars'', 
  the Klaus Tschira Stiftung, 
  the states of Baden-W\"urttemberg and Niedersachsen, 
  and by the Junta de Andaluc\'{\i}a.

We acknowledge the use of public TESS Alert data from pipelines at the TESS Science Office and at the TESS Science Processing Operations Center.
This research has made use of the Exoplanet Follow-up Observation Program website, which is operated by the California Institute of Technology, under contract with the National Aeronautics and Space Administration under the Exoplanet Exploration Program. Resources supporting this work were provided by the NASA High-End Computing (HEC) Program through the NASA Advanced Supercomputing (NAS) Division at Ames Research Center for the production of the SPOC data products.

We acknowledge financial support from
the European Research Council under the Horizon 2020 Framework Program via the ERC Advanced Grant Origins 83~24~28,
the Deutsche Forschungsgemeinschaft through projects 
    RE~281/32-1, 
    RE~1664/14-1,
    RE~2694/4-1,
    and RA714/14-1, PA525/18-1, PA525/19-1 within the Schwerpunkt SPP 1992,
the Agencia Estatal de Investigaci\'on of the Ministerio de Ciencia, Innovaci\'on y Universidades and the European FEDER/ERF funds through projects    
PGC2018-098153-B-C31, 
ESP2016-80435-C2-1-R,
ESP2016-80435-C2-2-R,
  AYA2016-79425-C3-1/2/3-P,     
  AYA2015-69350-C3-2-P, 
  RYC-2015-17697,          
  and BES-2017-082610,     
the Centre of Excellence ``Severo Ochoa'' and ``Mar\'{\i}a de Maeztu'' awards to the Instituto de Astrof\'isica de Canarias (SEV-2015-0548), Instituto de Astrof\'{\i}sica de Andaluc\'{\i}a (SEV-2017-0709), and Centro de Astrobiolog\'{\i}a (MDM-2017-0737), 
the European Union’s Horizon 2020 research and innovation program under the Marie Skłodowska-Curie grant
    713673,                
the Centre national d'études spatiales through grants
    PLATO                  
    and GOLF,              
the Czech Academy of Sciences through grant
    LTT20015,              
NASA through grants
    NNX17AF27G             
    and NNX17AG24G,
JSPS KAKENHI through grants
    JP18H01265
    and JP18H05439,
JST PRESTO through grant
    JPMJPR1775,
the Fundaci\'on Bancaria ``la Caixa'' through grant
    INPhINIT LCF/BQ/IN17/11620033,
and the Generalitat de Catalunya/CERCA programme.
NESSI was funded by the NASA Exoplanet Exploration Program and the NASA Ames Research Center and built at the Ames Research Center.
The authors are honored to be permitted to conduct observations on Iolkam Du'ag (Kitt Peak), a mountain within the Tohono O'odham Nation with particular significance to the Tohono O'odham people.
This work made use of observations from the LCOGT network and the following software:
\texttt{astrasens}, \texttt{AstroImageJ}, \texttt{Banzai}, \texttt{batman}, \texttt{caracal}, \texttt{emcee}, \texttt{juliet}, \texttt{serval}, \texttt{TESS Transit Finder}, \texttt{tpfplotter}, \texttt{Yabi}, and the python packages \texttt{astropy}, \texttt{lightkurve}, \texttt{matplotlib}, and \texttt{numpy}.
We thank the SuperWASP team and J.~Sanz-Forcada for sharing unpublished information with us.
Special thanks to Ismael Pessa for all their support through this work.
\end{acknowledgements}

\bibliographystyle{aa} 
\bibliography{biblio} 

\begin{appendix}

\section{Long tables}

\begin{table*}
    \centering
    \caption{Priors used for TOI-1235~b in the  joint fit with \texttt{juliet}.}
    \label{tab:priors}
    \begin{tabular}{lccr} 
        \hline
        \hline
        \noalign{\smallskip}
        Parameter$^a$  & Prior & Unit & Description \\
        \noalign{\smallskip}
        \hline
        \noalign{\smallskip}
        \multicolumn{4}{c}{\it Stellar parameters} \\
        \noalign{\smallskip}
        $\rho_\star$ & $\mathcal{N}(3.7,3.8)$ & g\,cm\,$^{-3}$ & Stellar density \\
        \noalign{\smallskip}
        \multicolumn{4}{c}{\it Planet parameters} \\
        \noalign{\smallskip}
        $P_{\rm b}$              & $\mathcal{U}(3,4)$           & d                    & Period of planet b \\
        $t_{0,b}$      & $\mathcal{U}(2458683,2458687)$     & d                    & Time of transit center of planet b \\
        $r_{1,b}$                & $\mathcal{U}(0,1)$           & \dots                & Parameterization for $p$ and $b$ \\
        $r_{2,b}$                & $\mathcal{U}(0,1)$           & \dots                & Parameterization for $p$ and $b$ \\
        $K_{b}$                  & $\mathcal{N}(0,100)$        & $\mathrm{m\,s^{-1}}$ & RV semi-amplitude of planet b \\
        $e_{b}$                & 0.0 (fixed)                    & \dots                & Orbital eccentricity of planet b \\
        $\omega_{b}$              & 90.0 (fixed)                    & deg              & Periastron angle of planet b \\
        \noalign{\smallskip}
        \multicolumn{4}{c}{\it Photometry parameters} \\
        \noalign{\smallskip}
        $D_{\mathrm{TESS}}$         & 1.0 (fixed)             & \dots     & Dilution factor for {\em TESS} Sectors 14, 20, and 21 \\
        $M_{\mathrm{TESS,S14}}$                & $\mathcal{N}(0,0.1)$    & \dots       & Relative flux offset for {\em TESS} Sector 14 \\
        $M_{\mathrm{TESS,S20}}$                & $\mathcal{N}(0,0.1)$    & \dots       & Relative flux offset for {\em TESS} Sector 20 \\
        $M_{\mathrm{TESS,S21}}$                & $\mathcal{N}(0,0.1)$    & \dots       & Relative flux offset for {\em TESS} Sector 21 \\
        $\sigma_{\mathrm{TESS,S14}}$           & $\mathcal{LU}(1,10^{4})$ & ppm       & Extra jitter term for {\em TESS} Sector 14 \\
        $\sigma_{\mathrm{TESS,S20}}$           & $\mathcal{LU}(1,10^{4})$ & ppm       & Extra jitter term for {\em TESS} Sector 20 \\
        $\sigma_{\mathrm{TESS,S21}}$           & $\mathcal{LU}(1,10^{4})$ & ppm       & Extra jitter term {\em TESS} Sector 21 \\
        $q_{1,\mathrm{TESS}}$        & $\mathcal{U}(0,1)$      & \dots     & Limb-darkening parameterization for {\em TESS} Sectors 14, 20, 21 \\
        $q_{2,\mathrm{TESS}}$        & $\mathcal{U}(0,1)$      & \dots     & Limb-darkening parameterization for {\em TESS} Sectors 14, 20, 21 \\
        $D_{\mathrm{LCO}}$                   & 1.0 (fixed)             & \dots     & Dilution factor for LCOGT \\
        $q_{1,\mathrm{LCO}}$                 & $\mathcal{U}(0,1)$      & \dots     & Limb-darkening parameterization for LCOGT \\
        $M_{\mathrm{LCO}}$                   & $\mathcal{N}(0,0.1)$    & \dots       & Relative flux offset for LCOGT \\
        $\sigma_{\mathrm{LCO}}$              & $\mathcal{LU}(1,10000)$ & ppm       & Extra jitter term for LCOGT \\
        $\theta_{0,\mathrm{LCO}}$              & $\mathcal{U}(-100,100)$ & \dots       & Extra jitter term for LCOGT \\
        $\theta_{1,\mathrm{LCO}}$              & $\mathcal{U}(-100,100)$ & \dots      & Extra jitter term for LCOGT \\
        \noalign{\smallskip}
        \multicolumn{4}{c}{\it RV parameters} \\
        \noalign{\smallskip}
        $\gamma_{\mathrm{HARPS-N}}$             & $\mathcal{N}(0,10)$     & $\mathrm{m\,s^{-1}}$ & RV zero-point for HARPS-N \\
        $\sigma_{\mathrm{HARPS-N}}$          & $\mathcal{LU}(0.01,10)$ & $\mathrm{m\,s^{-1}}$ & Extra jitter term for HARPS-N \\
        $\gamma_{\mathrm{CARMENES}}$            & $\mathcal{N}(0,10)$     & $\mathrm{m\,s^{-1}}$ & RV zero point for CARMENES \\
        $\sigma_{\mathrm{CARMENES}}$         & $\mathcal{LU}(0.01,10)$ & $\mathrm{m\,s^{-1}}$ & Extra jitter term for CARMENES \\
        \noalign{\smallskip}
        \noalign{\smallskip}
        \multicolumn{4}{c}{\it GP hyperparameters} \\
        \noalign{\smallskip}
        $\sigma_\mathrm{GP,RV}$                  & $\mathcal{LU}(10^{-10},100)$  & $\mathrm{m\,s^{-1}}$  & Amplitude of GP component for the RVs \\
        $\alpha_\mathrm{GP,RV}$                  & $\mathcal{LU}(10^{-10},100)$  & d$^{-2}$              & Inverse length-scale of GP exponential component for the RVs \\
        $\Gamma_\mathrm{GP,RV}$                  & $\mathcal{LU}(10^{-10},100)$  & \dots                  & Amplitude of GP sine-squared component for the RVs \\
        $P_\mathrm{rot;GP,RV}$                   & $\mathcal{U}(1,100)$          & d                     & Period of the GP quasi-periodic component for the RVs \\
        \noalign{\smallskip}
        \hline
    \end{tabular}
     \tablefoot{
        \tablefoottext{a}{The parameterization for $(p,b)$ was made with $(r_1,r_2)$ as in \citet{Espinoza18}. 
        The prior labels of $\mathcal{N}$, $\mathcal{U}$, and $\mathcal{LU}$ represent normal, uniform, and log-uniform distributions, respectively, where
        $\mathcal{N}(\mu,\sigma^2)$ is a normal distribution of the mean $\mu$ and variance $\sigma^2$ and $\mathcal{U}(a,b)$ and $\mathcal{LU}(a,b)$ are uniform and log-uniform distributions between $a$ and $b$}.}
\end{table*}

\begin{table*}
\centering
\small
\caption{Radial velocity measurements and spectroscopic activity indicators for TOI-1235 from optical spectra.} 
\label{tab:RV_Activity_all}
\begin{tabular}{l c c c c c c c c}
\hline
\hline
        \noalign{\smallskip}
        \multicolumn{9}{c}{CARMENES VIS} \\
        BJD & RV & CRX & dLW & H$\alpha$ & Ca~IRTa & TiO7050 & TiO8430 & TiO8860 \\
        (--2450000) & (m\,s$^{-1}$) & (m\,s$^{-1}$\,Np$^{-1}$) & (m$^{2}$\,s$^{-2}$) &\\
        \noalign{\smallskip}
        \hline
        \noalign{\smallskip}
8796.6533 & --1.2$\pm$2.2 & 9$\pm$19 & --30.4$\pm$3.8 & 0.6954$\pm$0.0021 & 0.5204$\pm$0.0022 & 0.8474$\pm$0.0015 & 0.8656$\pm$0.0027 & 0.9722$\pm$0.0028 \\
8807.7240 & 0.4$\pm$2.4 & --24$\pm$20 & --3.2$\pm$2.8 & 0.7004$\pm$0.0017 & 0.5221$\pm$0.0018 & 0.8487$\pm$0.0012 & 0.8694$\pm$0.0022 & 0.9684$\pm$0.0023 \\
8811.6588 & 6.3$\pm$2.8 & 37$\pm$25 & --22.0$\pm$4.1 & 0.6859$\pm$0.0027 & 0.5194$\pm$0.0029 & 0.8515$\pm$0.0020 & 0.8640$\pm$0.0036 & 0.9746$\pm$0.0035 \\
8815.7134 & --2.7$\pm$1.8 & 8$\pm$15 & --0.8$\pm$2.0 & 0.6907$\pm$0.0014 & 0.5294$\pm$0.0016 & 0.8494$\pm$0.0011 & 0.8665$\pm$0.0019 & 0.9742$\pm$0.0020 \\
8816.6576 & 3.0$\pm$1.6 & --14$\pm$13 & 5.5$\pm$2.0 & 0.6916$\pm$0.0013 & 0.5299$\pm$0.0015 & 0.8859$\pm$0.0010 & 0.8650$\pm$0.0018 & 0.9768$\pm$0.0018 \\
8817.7185 & 0.4$\pm$3.2 & 64$\pm$29 & --19.3$\pm$3.4 & 0.6973$\pm$0.0030 & 0.5223$\pm$0.0035 & 0.8379$\pm$0.0022 & 0.8542$\pm$0.0042 & 0.9759$\pm$0.0042 \\
8831.5414 & 5.4$\pm$2.8 & 0$\pm$23 & --4.6$\pm$3.3 & 0.6884$\pm$0.0020 & 0.5199$\pm$0.0022 & 0.8460$\pm$0.0015 & 0.8667$\pm$0.0027 & 0.9770$\pm$0.0027 \\
8832.6949 & 2.5$\pm$2.2 & --7$\pm$19 & 11.4$\pm$3.3 & 0.6893$\pm$0.0018 & 0.5223$\pm$0.0020 & 0.8486$\pm$0.0014 & 0.8629$\pm$0.0025 & 0.9782$\pm$0.0025 \\
8846.6694 & --4.2$\pm$1.6 & 16$\pm$14 & 6.3$\pm$1.6 & 0.6894$\pm$0.0011 & 0.5267$\pm$0.0013 & 0.8514$\pm$0.0009 & 0.8666$\pm$0.0016 & 0.9769$\pm$0.0017 \\
8848.7121 & 5.8$\pm$1.6 & 27$\pm$13 & 10.1$\pm$1.6 & 0.6963$\pm$0.0011 & 0.5200$\pm$0.0013 & 0.8489$\pm$0.0008 & 0.8640$\pm$0.0015 & 0.9765$\pm$0.0016 \\
8850.6431 & 2.8$\pm$1.4 & 12$\pm$13 & 1.5$\pm$1.9 & 0.6905$\pm$0.0012 & 0.5256$\pm$0.0014 & 0.8518$\pm$0.0009 & 0.8709$\pm$0.0017 & 0.9761$\pm$0.0017 \\
8852.6259 & 4.5$\pm$1.4 & 9$\pm$12 & 8.0$\pm$1.3 & 0.6989$\pm$0.0012 & 0.5191$\pm$0.0014 & 0.8502$\pm$0.0009 & 0.8687$\pm$0.0017 & 0.9812$\pm$0.0017 \\
8854.6620 & 6.0$\pm$2.1 & 27$\pm$14 & 5.5$\pm$1.8 & 0.6950$\pm$0.0011 & 0.5268$\pm$0.0013 & 0.8488$\pm$0.0009 & 0.8647$\pm$0.0016 & 0.9798$\pm$0.0016 \\
8855.6361 & 4.7$\pm$2.0 & --3$\pm$14 & 2.5$\pm$1.6 & 0.6864$\pm$0.0012 & 0.5254$\pm$0.0013 & 0.8499$\pm$0.0009 & 0.8678$\pm$0.0016 & 0.9754$\pm$0.0016 \\
8856.6278 & 0.0$\pm$1.5 & 13$\pm$11 & 6.6$\pm$1.7 & 0.6868$\pm$0.0011 & 0.5187$\pm$0.0013 & 0.8523$\pm$0.0009 & 0.8650$\pm$0.0015 & 0.9756$\pm$0.0016 \\
8857.6312 & --1.5$\pm$1.9 & 6$\pm$13 & 6.1$\pm$2.1 & 0.6952$\pm$0.0011 & 0.5283$\pm$0.0013 & 0.8491$\pm$0.0009 & 0.8666$\pm$0.0016 & 0.9832$\pm$0.0016 \\
8858.6017 & 6.7$\pm$2.0 & 15$\pm$16 & --2.3$\pm$2.5 & 0.6934$\pm$0.0017 & 0.5265$\pm$0.0018 & 0.8514$\pm$0.0013 & 0.8659$\pm$0.0022 & 0.9826$\pm$0.0023 \\
8860.6327 & --3.8$\pm$1.8 & --3$\pm$16 & 2.8$\pm$2.3 & 0.6859$\pm$0.0015 & 0.5227$\pm$0.0017 & 0.8538$\pm$0.0012 & 0.8665$\pm$0.0020 & 0.9769$\pm$0.0021 \\
8861.6279 & --3.8$\pm$1.9 & 26$\pm$15 & 3.2$\pm$2.3 & 0.6944$\pm$0.0014 & 0.5262$\pm$0.0015 & 0.8506$\pm$0.0010 & 0.8665$\pm$0.0019 & 0.9794$\pm$0.0019 \\
8862.6304 & --0.2$\pm$1.8 & 27$\pm$14 & --0.7$\pm$1.9 & 0.6947$\pm$0.0014 & 0.5285$\pm$0.0016 & 0.8509$\pm$0.0011 & 0.8693$\pm$0.0019 & 0.9768$\pm$0.0020 \\
8863.6852 & --1.4$\pm$3.0 & --48$\pm$27 & 2.7$\pm$3.5 & 0.6932$\pm$0.0021 & 0.5312$\pm$0.0023 & 0.8470$\pm$0.0016 & 0.8613$\pm$0.0028 & 0.9778$\pm$0.0029 \\
8864.6148 & --3.8$\pm$2.1 & --11$\pm$16 & --6.6$\pm$2.4 & 0.6862$\pm$0.0016 & 0.5217$\pm$0.0018 & 0.8507$\pm$0.0012 & 0.8662$\pm$0.0021 & 0.9758$\pm$0.0021 \\
8865.6228 & 0.6$\pm$2.5 & --33$\pm$23 & --5.5$\pm$2.6 & 0.6967$\pm$0.0019 & 0.5277$\pm$0.0021 & 0.8477$\pm$0.0014 & 0.8648$\pm$0.0025 & 0.9864$\pm$0.0026 \\
8866.6362 & 2.0$\pm$3.4 & --23$\pm$28 & --4.4$\pm$3.5 & 0.6948$\pm$0.0028 & 0.5287$\pm$0.0030 & 0.8479$\pm$0.0020 & 0.8626$\pm$0.0036 & 0.9867$\pm$0.0036 \\
8877.5779 & --2.4 $\pm$2.0 & --10$\pm$13 & --4.4$\pm$2.2 & 0.6927$\pm$0.0011 & 0.5219$\pm$0.0012 & 0.8467$\pm$0.0008 & 0.8682$\pm$0.0015 & 0.9839$\pm$0.0015 \\
8881.5843 & --0.6$\pm$1.6 & 13$\pm$15 & --8.0$\pm$1.9 & 0.6875$\pm$0.0012 & 0.5185$\pm$0.0013 & 0.8471$\pm$0.0009 & 0.8670$\pm$0.0016 & 0.9821$\pm$0.0017 \\
8882.5742 & 0.3$\pm$1.6 & 2$\pm$14 & --19.2$\pm$2.2 & 0.6948$\pm$0.0014 & 0.5208$\pm$0.0015 & 0.8493$\pm$0.0010 & 0.8687$\pm$0.0018 & 0.9793$\pm$0.0019 \\
8883.5713 & --5.9$\pm$1.6 & 1$\pm$13 & --9.5$\pm$1.7 & 0.6954$\pm$0.0012 & 0.5210$\pm$0.0014 & 0.8485$\pm$0.0009 & 0.8679$\pm$0.0017 & 0.9821$\pm$0.0017 \\
8884.5713 & --5.5$\pm$1.2 & 2.0$\pm$8.8 & --4.9$\pm$1.6 & 0.6912$\pm$0.0011 & 0.5151$\pm$0.0012 & 0.8473$\pm$0.0008 & 0.8675$\pm$0.0014 & 0.9756$\pm$0.0015 \\
8885.5794 & 1.2$\pm$1.7 & --8$\pm$13 & --5.9$\pm$1.8 & 0.6859$\pm$0.0013 & 0.5209$\pm$0.0014 & 0.8468$\pm$0.0010 & 0.8701$\pm$0.0017 & 0.9804$\pm$0.0018 \\
8887.5650 & --9.7$\pm$3.2 & 17$\pm$27 & --9.4$\pm$3.7 & 0.6919$\pm$0.0027 & 0.5260$\pm$0.0028 & 0.8484$\pm$0.0020 & 0.8684$\pm$0.0035 & 0.9772$\pm$0.0035 \\
8888.7326 & 2.2$\pm$3.7 & 12$\pm$34 & 11.0$\pm$4.6 & 0.6912$\pm$0.0032 & 0.5210$\pm$0.0034 & 0.8472$\pm$0.0024 & 0.8611$\pm$0.0043 & 0.9758$\pm$0.0041 \\
8890.5100 & --0.1$\pm$2.8 & --37$\pm$24 & 2.3$\pm$2.0 & 0.6939$\pm$0.0016 & 0.5295$\pm$0.0018 & 0.8487$\pm$0.0012 & 0.8651$\pm$0.0022 & 0.9798$\pm$0.0022 \\
8890.5332 & --1.5$\pm$1.8 & --19$\pm$13 & 2.2$\pm$1.8 & 0.6884$\pm$0.0014 & 0.5268$\pm$0.0016 & 0.8500$\pm$0.0011 & 0.8668$\pm$0.0019 & 0.9816$\pm$0.0020 \\
8891.5446 & --2.6$\pm$1.5 & 4$\pm$12 & 6.0$\pm$2.4 & 0.7054$\pm$0.0012 & 0.5305$\pm$0.0014 & 0.8525$\pm$0.0009 & 0.8677$\pm$0.0016 & 0.9872$\pm$0.0017 \\
8893.5107 & 4.1$\pm$1.6 & 4$\pm$13 & 7.9$\pm$2.2 & 0.6982$\pm$0.0013 & 0.5270$\pm$0.0014 & 0.8500$\pm$0.0010 & 0.8705$\pm$0.0017 & 0.9800$\pm$0.0018 \\
8894.5328 & --1.0$\pm$1.9 & --24$\pm$14 & 8.1$\pm$1.4 & 0.6911$\pm$0.0010 & 0.5297$\pm$0.0011 & 0.8522$\pm$0.0008 & 0.8678$\pm$0.0014 & 0.9802$\pm$0.0015 \\
8895.5580 & 0.6$\pm$2.0 & --52$\pm$14 & --4.9$\pm$2.2 & 0.6938$\pm$0.0013 & 0.5265$\pm$0.0014 & 0.8506$\pm$0.0010 & 0.8706$\pm$0.0018 & 0.9865$\pm$0.0018 \\
8896.5272 & 3.0$\pm$1.4 & --8$\pm$12 & 0.3$\pm$1.9 & 0.6989$\pm$0.0011 & 0.5233$\pm$0.0012 & 0.8510$\pm$0.0008 & 0.8682$\pm$0.0015 & 0.9784$\pm$0.0015 \\
8897.5334 & --3.0$\pm$1.4 & --15$\pm$11 & 1.5$\pm$1.5 & 0.6990$\pm$0.0012 & 0.5270$\pm$0.0013 & 0.8489$\pm$0.0009 & 0.8642$\pm$0.0016 & 0.9877$\pm$0.0017 \\
\hline
        \noalign{\smallskip}
        \multicolumn{9}{c}{HARPS-N} \\
        BJD & RV & CRX & dLW & H$\alpha$ & $S_{\rm MWO}$ & $\log{R'_{\rm HK}}$ \\
        (--2450000) & (m\,s$^{-1}$) & (m\,s$^{-1}$\,Np$^{-1}$) & (m$^{2}$\,s$^{-2}$) &\\
        \noalign{\smallskip}
        \hline
        \noalign{\smallskip}
8862.5810 & 6.6$\pm$2.0 & 11$\pm$17 & --12.7$\pm$3.5 & 0.7144$\pm$0.0026 & 0.967$\pm$0.043 & --4.748$\pm$0.045 \\
8862.7100 & 2.0$\pm$1.0 & --2.2$\pm$7.5 & --25.8$\pm$2.0 & 0.7125$\pm$0.0013 & 0.991$\pm$0.014 & --4.735$\pm$0.038 \\
8863.6284 & 0.24$\pm$0.60 & 2.1$\pm$4.8 & --30.4$\pm$1.3 & 0.7154$\pm$0.0010 & 1.015$\pm$0.008 & --4.722$\pm$0.037\\
8863.7378 & --1.32$\pm$0.72 & 4.7$\pm$5.7 & --30.4$\pm$1.3 & 0.7147$\pm$0.0009 & 1.005$\pm$0.009 & --4.727$\pm$0.037\\
8864.6070 & 5.2$\pm$1.0 & 19.5$\pm$7.6 & --28.4$\pm$1.7 & 0.7144$\pm$0.0013 & 1.000$\pm$0.012 & --4.730$\pm$0.038 \\
8864.7169 & 5.91$\pm$0.99 & 7.6$\pm$7.8 & --30.4$\pm$1.6 & 0.7242$\pm$0.0013 & 1.040$\pm$0.014 & --4.709$\pm$0.038 \\
8865.5889 & 8.28$\pm$0.70 & --0.9$\pm$5.7 & --29.8$\pm$1.4 & 0.7156$\pm$0.0012 & 1.027$\pm$0.010 & --4.716$\pm$0.037 \\
8865.7198 & 4.97$\pm$0.76 & 8.8$\pm$6.0 & --29.0$\pm$1.3 & 0.7134$\pm$0.0011 & 1.009$\pm$0.008 & --4.725$\pm$0.037\\
8869.6175 & 10.75$\pm$0.98 & 4.5$\pm$7.8 & --25.9$\pm$2.2 & 0.7131$\pm$0.0015 & 0.986$\pm$0.021 & --4.738$\pm$0.039 \\
8869.7536 & 10.22$\pm$0.89 & --4.2$\pm$6.9 & --26.1$\pm$1.5 & 0.7152$\pm$0.0009 & 1.073$\pm$0.011 & --4.692$\pm$0.037 \\
8870.6093 & 6.0$\pm$1.6 & --22$\pm$13 & --22.4$\pm$2.4 & 0.7120$\pm$0.0018 & 1.031$\pm$0.030 & --4.713$\pm$0.040 \\
8870.6978 & 6.0$\pm$1.6 & --21$\pm$13 & --22.4$\pm$2.9 & 0.7087$\pm$0.0020 & 1.017$\pm$0.033 & --4.721$\pm$0.041 \\
8895.4528 & 11.9$\pm$1.9 & 7$\pm$15 & --27.5$\pm$2.8 & 0.7159$\pm$0.0024 & 0.999$\pm$0.042 & --4.731$\pm$0.044 \\
8896.5214 & 11.37$\pm$0.95 & --2.2$\pm$7.5 & --31.2$\pm$1.9 & 0.7186$\pm$0.0013 & 0.985$\pm$0.016 & --4.738$\pm$0.038 \\
8896.6331 & 12.1$\pm$1.2 & --2.2$\pm$9.6 & --31.7$\pm$1.8 & 0.7155$\pm$0.0012 & 1.021$\pm$0.014 & --4.719$\pm$0.038 \\
8897.6418 & 4.6$\pm$1.1 & --14.0$\pm$8.4 & --34.1$\pm$1.5 & 0.7166$\pm$0.0012 & 0.978$\pm$0.012 & --4.742$\pm$0.038 \\
8898.5249 & 6.64$\pm$0.87 & --16.4$\pm$6.4 & --35.4$\pm$1.3 & 0.7170$\pm$0.0010 & 1.009$\pm$0.010 & --4.725$\pm$0.037 \\
8898.6937 & 2.4$\pm$1.0 & --8.2$\pm$8.2 & --35.0$\pm$1.8 & 0.7234$\pm$0.0015 & 0.980$\pm$0.019 & --4.741$\pm$0.039 \\
8905.5116 & --0.8$\pm$1.4 & 2$\pm$11 & --36.8$\pm$2.6 & 0.7242$\pm$0.0018 & 0.989$\pm$0.025 & --4.736$\pm$0.040 \\
8905.6346 & --0.5$\pm$1.6 & --8$\pm$13 & --36.3$\pm$2.3 & 0.7185$\pm$0.0020 & 0.942$\pm$0.026 & --4.763$\pm$0.041 \\
8925.5936 & --5.1$\pm$1.5 & 3$\pm$12 & --27.1$\pm$2.3 & 0.7183$\pm$0.0020 & 1.033$\pm$0.031 & --4.712$\pm$0.040 \\
        \hline
\end{tabular}
\end{table*}

\begin{table*}
\centering
\small
\caption{Radial velocity measurements and spectroscopic activity indicators for TOI-1235 from NIR spectra.} 
\label{tab:RV_NIR}
\begin{tabular}{l c c c }
\hline
\hline
        \noalign{\smallskip}
        \multicolumn{4}{c}{CARMENES NIR} \\
        BJD & RV & CRX & dLW \\
        (--2450000) & (m\,s$^{-1}$) & (m\,s$^{-1}$\,Np$^{-1}$) & (m$^{2}$\,s$^{-2}$)\\
        \noalign{\smallskip}
        \hline
        \noalign{\smallskip}
8796.6534 & --3.62 $\pm$ 8.7 & --15.51 $\pm$ 43 & --9.56 $\pm$ 16 \\
8807.7242 & 2.7 $\pm$ 11.5 & 35.98 $\pm$ 51 & 14.64 $\pm$ 11 \\
8815.7113 & --5.57 $\pm$ 7.0 & 59.31 $\pm$ 32 & 20.36 $\pm$ 6.3 \\
8816.6574 & --12.1 $\pm$ 4.3 & 50.87 $\pm$ 17 & 26.59 $\pm$ 12 \\
8817.7193 & --28.31 $\pm$ 12.5 & 91.45 $\pm$ 58 & 2.89 $\pm$ 24 \\
8831.5418 & 14.98 $\pm$ 10.8 & 25.22 $\pm$ 51 & 20.13 $\pm$ 16 \\
8832.6942 & --9.92 $\pm$ 9.5 & 102.59 $\pm$ 39 & 32.11 $\pm$ 11 \\
8846.6695 & 16.88 $\pm$ 14.2 & --169.25 $\pm$ 98 & --55.07 $\pm$ 11 \\
8848.7123 & 9.06 $\pm$ 7.1 & 21.37 $\pm$ 37 & 19.54 $\pm$ 5.2 \\
8850.6443 & --7.15 $\pm$ 7.5 & 41.75 $\pm$ 37 & 10.7 $\pm$ 6.1 \\
8852.6275 & --3.75 $\pm$ 6.2 & --18.49 $\pm$ 31 & 25.48 $\pm$ 8.5 \\
8854.6613 & 3.59 $\pm$ 8.7 & --23.5 $\pm$ 40 & 12.99 $\pm$ 9.2 \\
8855.6350 & --25.78 $\pm$ 14.4 & --127.78 $\pm$ 99 & --10.78 $\pm$ 13\\
8856.6274 & --16.5 $\pm$ 5.1 & 37.01 $\pm$ 23 & 28.13 $\pm$ 8.2 \\
8857.6309 & 0.07 $\pm$ 4.8 & 12.28 $\pm$ 22 & 9.51 $\pm$ 7.9 \\
8858.6034 & --12.53 $\pm$ 5.6 & --13.31 $\pm$ 23 & --8.3 $\pm$ 10 \\
8860.6337 & --7.52 $\pm$ 7.3 & 95.19 $\pm$ 27 & 24.34 $\pm$ 11 \\
8861.6289 & --3.69 $\pm$ 5.5 & 28.43 $\pm$ 26 & 19.61 $\pm$ 9.8 \\
8862.6275 & 10.29 $\pm$ 8.1 & 84.9 $\pm$ 36 & 31.28 $\pm$ 9.3 \\
8863.6678 & --27.79 $\pm$ 25.7 & 56.84 $\pm$ 127 & 63.0 $\pm$ 29 \\
8863.6864 & --10.75 $\pm$ 13.1 & 14.07 $\pm$ 65 & 27.92 $\pm$ 11 \\
8864.6146 & --9.35 $\pm$ 8.6 & 108.24 $\pm$ 34 & 44.67 $\pm$ 8.2 \\
8865.6222 & 6.09 $\pm$ 9.3 & 82.43 $\pm$ 42 & 0.45 $\pm$ 8.3 \\
8866.6361 & --24.99 $\pm$ 19.1 & 46.87 $\pm$ 108 & --107.27 $\pm$ 34\\
8877.5768 & --11.17 $\pm$ 5.6 & 7.22 $\pm$ 27 & 14.44 $\pm$ 6.4\\
8881.5859 & --7.27 $\pm$ 5.8 & 61.08 $\pm$ 25 & 3.89 $\pm$ 6.7\\
8882.5755 & --5.17 $\pm$ 5.0 & --12.44 $\pm$ 24 & --2.18 $\pm$ 6.8\\
8883.5726 & --6.14 $\pm$ 6.6 & 38.85 $\pm$ 32 & --3.2 $\pm$ 6.8\\
8884.5719 & --7.81 $\pm$ 4.9 & 24.28 $\pm$ 23 & --4.16 $\pm$ 8.1\\
8885.5807 & --12.04 $\pm$ 7.0 & 29.82 $\pm$ 34 & 19.68 $\pm$ 9.0 \\
8887.5639 & --5.08 $\pm$ 10.8 & --104.73 $\pm$ 46 & --15.27 $\pm$ 25\\
8888.7295 & --12.98 $\pm$ 17.3 & 34.81 $\pm$ 85 & --27.62 $\pm$ 28 \\
8890.5100 & 1.13 $\pm$ 15.4 & --5.12 $\pm$ 113 & --34.55 $\pm$ 18 \\
8890.5349 & --8.16 $\pm$ 7.1 & --20.73 $\pm$ 38 & 18.13 $\pm$ 10 \\
8891.5421 & --1.74 $\pm$ 4.8 & --12.32 $\pm$ 24 & 10.26 $\pm$ 8.6 \\
8893.5073 & 9.0 $\pm$ 7.6 & 47.15 $\pm$ 45 & 38.66 $\pm$ 14 \\
8894.5324 & --8.05 $\pm$ 5.3 & 55.52 $\pm$ 24 & 5.35 $\pm$ 6.9 \\
8895.5593 & 4.62 $\pm$ 6.1 & --33.6 $\pm$ 30 & 3.3 $\pm$ 7.3 \\
8896.5263 & 0.14 $\pm$ 5.1 & --11.55 $\pm$ 27 & 32.04 $\pm$ 6.6 \\
8897.5325 & 5.19 $\pm$ 13.7 & --22.3 $\pm$ 100 & --41.89 $\pm$ 19 \\
8903.4969 & --4.6 $\pm$ 10.7 & 139.99 $\pm$ 63 & 36.28 $\pm$ 16 \\
8904.4842 & --11.51 $\pm$ 6.1 & --19.26 $\pm$ 29 & 6.46 $\pm$ 10 \\
\hline
        \noalign{\smallskip}
        \multicolumn{2}{c}{iSHELL} \\
        BJD & RV \\
        (--2450000) & (m\,s$^{-1}$) \\
        \noalign{\smallskip}
        \hline
        \noalign{\smallskip}
8874.1303 & --0.71 $\pm$ 9.1 \\
8875.1161 & 6.61 $\pm$ 11.1 \\
8895.0886 & 2.6 $\pm$ 4.0 \\
8899.0817 & 13.07 $\pm$ 5.2 \\
8901.0644 & --4.16 $\pm$ 6.1 \\
        \hline
\end{tabular}
\end{table*}

\end{appendix}

\end{document}